\documentclass[10pt, aps, prd, superscriptaddress, nofootinbib, showpacs, twocolumn]{revtex4-2}
\usepackage[english]{babel}
\usepackage{amsmath, amsfonts, amsthm, amssymb, mathrsfs}
\usepackage[all]{xy}
\usepackage{bm}
\usepackage{stmaryrd}
\usepackage{graphicx}
\usepackage[colorlinks]{hyperref}
\usepackage{color}
\usepackage{subcaption}
\usepackage{amscd}

\usepackage{pgfplots, grffile, tikz}
\pgfplotsset{compat=newest}
\usetikzlibrary{plotmarks, arrows.meta}
\usepgfplotslibrary{patchplots}

\usepackage{dsfont}

\numberwithin{equation}{section}

\DeclareMathOperator{\tr}{tr}
\DeclareMathOperator{\Tr}{Tr}

\renewcommand{\Re}{\mathop{\mathrm{Re}}\nolimits}

\newcommand{\calE}{\mathcal E}
\newcommand{\calG}{\mathcal G}
\newcommand{\calK}{\mathcal K}
\newcommand{\calL}{\mathcal L}
\newcommand{\calM}{\mathcal M}
\newcommand{\calP}{\mathcal P}
\newcommand{\calR}{\mathcal R}

\newcommand{\bbB}{\mathbb B}
\newcommand{\bbC}{\mathbb C}
\newcommand{\bbW}{\mathbb W}
\newcommand{\bbL}{\mathbb L}
\newcommand{\bbM}{\mathbb M}

\newcommand{\bbR}{\mathbb R}
\newcommand{\bbZ}{\mathbb Z}

\newcommand{\frakM}{\mathfrak{M}}
\newcommand{\frakL}{\mathfrak{L}}

\newcommand{\lp}{\left(}
\newcommand{\rp}{\right)}
\newcommand{\lb}{\left[}
\newcommand{\rb}{\right]}
\newcommand\tln[2]{\substack{#1\\#2}}
\renewcommand{\a}{\alpha}
\newcommand{\m}{\mu}
\newcommand{\n}{\nu}
\newcommand{\s}{\sigma}
\newcommand{\nb}{\nabla}

\newcommand{\h}{\hat}
\renewcommand{\t}{\widetilde}

\begin{document}

\title{Functorial properties of Schwinger--DeWitt expansion and Mellin--Barnes representation}

\author{A. O. Barvinsky}
\email{barvin@td.lpi.ru}
\affiliation{Theory Department, Lebedev Physics Institute, Leninsky Prospect 53, Moscow 119991, Russia}
\affiliation{Institute for Theoretical and Mathematical Physics, Moscow State University, Leninskie Gory, GSP-1, Moscow, 119991, Russia}

\author{A. E. Kalugin}
\email{kalugin.ae@phystech.edu}
\affiliation{Theory Department, Lebedev Physics Institute, Leninsky Prospect 53, Moscow 119991, Russia}

\author{W. Wachowski}
\email{vladvakh@gmail.com}
\affiliation{Theory Department, Lebedev Physics Institute, Leninsky Prospect 53, Moscow 119991, Russia}

\begin{abstract}
We consider integral kernels for functions $f(\hat F)$ of a minimal second-order differential operator $\hat F(\nabla)$ on a curved spacetime. We show that they can be expanded in a functional series, analogous to the DeWitt expansion for the heat kernel, by integrating the latter term-by-term. This procedure leads to a separation of two types of data: all information about the bundle geometry and the operator $\hat F(\nabla)$ is still contained in the standard HaMiDeW coefficients $\hat a_k[F | x,x']$ (we call this property ``off-diagonal functoriality''), while information about the function $f$ is encoded in some new scalar functions $\bbB_\alpha[f | \sigma]$ and $\bbW_\alpha[f | \sigma, m^2]$, which we call basis and complete massive kernels, respectively. These objects are calculated for operator functions of the form $\exp(-\tau\hat F^\nu)/(\hat F^\mu + \lambda)$ as multiple Mellin--Barnes integrals. The article also discusses subtle issues such as the validity of the term-by-term integration, the regularization of IR divergent integrals, and the physical interpretation of the resulting expansions.
\end{abstract}

\maketitle
\tableofcontents

\section{Introduction} \label{sec:introduction}

The heat kernel method is an important tool in the arsenal of modern mathematical physics~\cite{DeWitt1965, Barvinsky1985, Gilkey1975, Gilkey1995}. Let us briefly remind the reader the basics of it while simultaneously fixing our conventions. Let $\hat F(\nabla)$ be a non-negatively definite (pseudo)differential operator acting on sections of a vector bundle\footnote{Hats everywhere indicate the matrix nature of the objects under consideration, for example, $\hat 1=\delta^A_B$, where $A, B$ are indices in fibers (they run over the set of all fields of the theory and can have both internal, spacetime, and spinor components). Unless otherwise stated, we will assume that when an operator acts on a two-point function, it always acts with respect to its first argument.} over a Riemannian manifold $\calM$ with metric $g_{ab}$. Then the integral kernel of the function of $\h F(\nb)$ is
\begin{equation} \label{pdo_kernel}
\hat K[f(F) | x,x'] = f(\hat F)\, \delta(x,x'),
\end{equation}
where $\delta(x,x')$ is the delta function. It is a two-point (i.e., depending on two points $x,x'\in\calM$) matrix-valued function. Then the action of an operator $f(\h F)$ on the section of the bundle $\bm{\varphi} = \varphi^A$ is represented by an integral convolution with the corresponding kernel:
\begin{equation}
f(\hat F)\,\bm{\varphi}(x) = \int\limits_\calM d^dx'\, \hat K[\,f(F) | x, x'] \,\bm{\varphi}(x').
\end{equation}

In particular, the heat kernel is the integral kernel of the operator exponential
\begin{equation} \label{HeatKernelDef}
\hat K_F(\tau | x,x') = \hat K[e^{-\tau F} | x, x'],
\end{equation}
where the additional parameter $\tau$ is referred to as proper time.

The heat kernel finds application in many areas of both pure mathematics and theoretical physics. In particular, its applications in quantum field theory (QFT) on a curved background are based on the fact that if $\calM$ is a physical spacetime, and $\hat F[\bm{\Phi} | \nabla]$ is the wave operator for some field theory (depending on nontrivial background fields $\bm{\Phi} = \Phi^A$), then the one-loop quantum effective action of this theory is given by
\begin{equation} \label{effective_action}
\varGamma_\text{1-loop}[\bm{\Phi}] = \frac12 \Tr\ln\hat F = -\frac12 \int\limits_0^\infty \frac{d\tau}\tau\, \Tr e^{-\tau\hat F},
\end{equation}
where the functional trace $\Tr$ is defined as
\begin{equation} \label{FuncTrDef}
\Tr f(\hat F) = \int\limits_\calM d^dx\; \tr\hat K[f(F)| x, x], \quad \tr\hat K = K^A_A.
\end{equation}

The efficiency of the heat kernel method is based on the fact that for a second order minimal operator
\begin{equation} \label{minimal}
\hat F(\nabla) = -\hat1 \Box + \hat P,
\end{equation}
where $\Box = -g^{ab}\nabla_a\nabla_b$ is the covariant Laplacian\footnote{We work in the Euclidean version of the theory related to the physical Lorentzian signature spacetime by Wick rotation.} and $\hat P(x)$ is a matrix depending on the point $x$ (a potential term), there is a remarkable DeWitt's expansion for the heat kernel $\hat K_F(\tau | x,x')$ in the form of a functional series~\cite{DeWitt1965}\footnote{\label{footnote1}A generalization of this expansion to the case of higher-order minimal operators $\hat F(\nabla) = \hat1 (-\Box)^N + \hat P(\nabla)$ was constructed in \cite{Wach2, Wach3, BKW2024}, and in the paper \cite{Barvinsky25} it was shown how to obtain the corresponding expansion for causal (non-minimal) operators.},
\begin{equation} \label{HeatKernelExpansion}
\hat K_F(\tau | x,x') = \sum\limits_{k=0}^\infty B_{\tfrac{d}{2} - k}(\tau,\sigma) \cdot \hat a_k[F | x,x'].
\end{equation}
Here the scalar function
\begin{equation} \label{InitialKernel}
B_\alpha(\tau ,\sigma) = \frac{\tau^{-\alpha}}{(4\pi)^{d/2}} \exp\left(-\frac{\sigma}{2\tau}\right)
\end{equation}
depends neither on the bundle geometry nor on the potential term $\hat P$, and its dependence on the points $x$ and $x'$ is only implicit, through its dependence on the Synge world function $\sigma(x,x')$ (i.e., one half of the square of the geodesic distance between $x$ and $x'$). Conversely, the off-diagonal heat kernel coefficients $\hat a_k[F | x,x']$ (which we will also call HaMiDeW in honor of Hadamard, Minakshisundaram, and DeWitt~\cite{Gibbons}) are matrix-valued two-point functions which do not depend on the proper time $\tau$, but contain information about the bundle geometry and the potential term $\hat P$.\footnote{It is always assumed that $x$ and $x'$ are sufficiently close to each other so that all quantities under consideration are well-defined. In our notations the Pauli--van Vleck--Morette determinant $\Delta^{1/2}(x,x')$ is absorbed into coefficients $\hat a_k[F | x,x']$ of the expansion \eqref{HeatKernelExpansion}.}

The coincidence limits of the HaMiDeW coefficients and their covariant derivatives can be found from the chain of recurrent equations for $\hat a_k[F | x,x']$ as some combinations of the background field curvatures $\mathfrak{R} \in \{R_a{}_{bcd}, \hat\calR_{ab}, \hat P\}$ and their derivatives \cite{DeWitt1965,Barvinsky1985}, which can be symbolically represented as
\begin{equation} \label{BackgroundDimensionEq}
\nabla^n \hat a_k[F | x,x']\,\Big|_{\,x=x'} = \sum\limits_{2m+l = 2k+n} \nabla^l \mathfrak{R}^m.
\end{equation}
Then substituting \eqref{HeatKernelExpansion} into the formula \eqref{effective_action} yields a local expansion of the one-loop effective action $\varGamma_\text{1-loop}[\bm{\Phi}]$ in a series in increasing powers of the background dimension, which underlies effective field theory.

In many applications there is a need to obtain an off-diagonal (i.e., when $x\ne x'$) expansion similar to the DeWitt one \eqref{HeatKernelExpansion} not only for the operator exponential $\exp(-\tau\hat F)$, but also of more {complicated} operator functions $f(\hat F)$.

The physical application that motivates us most is the need to study models with a \emph{non-minimal} wave operator $\hat H(\nabla)$. Non-minimality of the operator means that its principal term can no longer be represented as a power of the d'Alembertian $\Box$. Such models, in particular the Ho\v{r}ava--Lifshitz type, are getting more and more frequently encountered in high energy physics and quantum gravity \cite{Barvinsky17, Barvinsky172, BarvinskyKurov2022, BarvinskyKurov2023}. They are either too complicated to be handled by the usual Feynman diagrammatic technique or impossible to treat by the original Schwinger--DeWitt method. An important special case are \emph{causal} models, i.e., those in which the characteristic surfaces of physical modes (in the Lorentzian version of the theory) are null surfaces in the metric $g_{ab}$. As shown in \cite{Barvinsky25, BarvinskyKalugin2024}, calculating the heat kernel $\exp(-\tau\hat H)$ and the effective action \eqref{effective_action} for such models can be reduced to finding expansions for operator functions of a more complex form
\begin{equation} \label{Operator}
f(\hat F)=\frac{e^{\textstyle -\tau \hat F^\nu}}{\vphantom{L^L}\hat F^\mu},
\end{equation}
where the operator $\hat F(\nabla)$ \eqref{minimal} is the second-order minimal, and the powers $\mu$ and $\nu$ are determined by the order of the operator $\hat H(\nabla)$ and the spin-tensor content of the model. Therefore, the key problem that we solve further in the proposed paper is the construction of expansions similar to DeWitt's \eqref{HeatKernelExpansion} for kernels of operator functions of the form \eqref{Operator}.

On the other hand, in mathematics, expansions for functional traces of the form $\Tr\big[Q(\hat F)\,\exp(-\tau\hat F)\big]$, where $Q(\h F)$ is a polynomial, with remarkable functorial properties are often considered~\cite{Seeley, Gilkey1975, Gilkey1995, GilkeyFegan}. However, since computing the functional trace \eqref{FuncTrDef} requires going to the diagonal of the kernel, such objects require subtle refinement using various techniques such as $\zeta$-functional regularization and analytic continuation. Therefore, naturally there is a necessity to obtain them in a much simpler way, as coincidence limits of the corresponding off-diagonal objects, thus employing what one could call \textit{off-diagonal functoriality}.

The analogue of the expansion~\eqref{HeatKernelExpansion} for a generic operator function $f(\hat F)$ can be obtained by using the linear integral transform $\frakL_f$ applicable to any admissible function $\varphi(\tau)$ of the proper time parameter $\tau$,
\begin{equation} \label{defLf}
\frakL_f\,\varphi = \int\limits_0^\infty d\tau\, f^*(\tau)\,\varphi(\tau),
\end{equation}
where $f^*(\tau)$ is the inverse Laplace transform of $f(\lambda)$
\begin{equation} \label{InvLaplaceTransform}
f^*(\tau) = \int\limits_C \frac{d\lambda}{2\pi i}\; f(\lambda)\, e^{\tau\lambda}.
\end{equation}
When applied to $\varphi(\tau)=\exp(-\tau\hat F)$, the transform \eqref{defLf} gives the needed operator function $f(\hat F)$ (see formula \eqref{GenIntTransform} below)
\begin{equation} \label{frakLtransformEq}
f(\hat F) = \frakL_f\, e^{-\tau\hat F} = \int\limits_0^\infty d\tau\, f^*(\tau)\,e^{-\tau\hat F}.
\end{equation}
This representation can be understood as the direct Laplace transform from the complex variable $\tau$ to the operator variable $\hat F$.

Rewriting this transformation on the kernel level, then substituting the DeWitt expansion \eqref{HeatKernelExpansion} into it, and finally taking the integrals \emph{term-by-term} we obtain the following generalization of the expansion \eqref{HeatKernelExpansion}:
\begin{equation} \label{general_kernel_series_rep}
\hat K\!\big[f(F) \big| x,x' \big] = \sum\limits_{k=0}^\infty \bbB_{\tfrac{d}{2}-k}[f | \sigma] \cdot \hat a_k[F | x,x'],
\end{equation}
where $\hat a_k[ F | x,x']$ are exactly the same off-diagonal HaMiDeW coefficients as in the expansion \eqref{HeatKernelExpansion}, and $\bbB_\alpha[f | \sigma]$ is an analytic function of the parameters $\sigma$ and $\alpha$, obtained from $B_\alpha(\tau,\sigma)$ \eqref{InitialKernel} by means the above-mentioned integral transform $\frakL_f$
\begin{equation} \label{frakLtransformEq2}
\bbB_\alpha[f | \sigma] = \frakL_f B_\alpha(\tau,\sigma).
\end{equation}

In other words, this procedure explicitly accounts for the separation of two different types of data into two different objects. In the expansion \eqref{general_kernel_series_rep}, all information about the bundle geometry and the original operator $\hat F(\nabla)$ is still encoded in the familiar to us HaMiDeW coefficients $\hat a_k[ F | x,x']$. While the functions $\bbB_\alpha\![f | \sigma]$, which we will call \emph{basis kernels}, do not depend neither on the geometry, nor on the specific form of the operator $\hat F(\nabla)$, but are determined exclusively by the function $f$.

The remarkable property that for any function $f$ the expansion \eqref{general_kernel_series_rep} includes the same HaMiDeW coefficients $\hat a_k[ F | x,x']$ is an off-diagonal generalization of the property that Gilkey and others call ``functoriality'' \cite{GilkeyFegan, Gilkey1975, Gilkey1995}. Therefore, we call this property \emph{off-diagonal functoriality}. Off-diagonality is a new and key ingredient here, providing convenience and flexibility of the approach. This is due to the fact that, being essentially a form of regularization by separating points, it allows one to avoid the singularities that arise in the coincidence limit $x=x'$, and to use the powerful apparatus of integral transforms more effectively.

As we explained in \cite{BKWLetter}, the idea of term-by-term integration of the DeWitt expansion \eqref{HeatKernelExpansion} presented above encounters an obvious objection: according to the well-known Fubini--Tonelli theorem, changing the order of summation/integration is only possible if all intermediate sums/integrals are absolutely convergent. In our case, this condition is clearly not satisfied, since, firstly, the DeWitt series \eqref{HeatKernelExpansion} is not convergent, but only asymptotic (divergent), and, secondly, the integrals in \eqref{frakLtransformEq2} also diverge at the IR limit $\tau=\infty$.

Despite this, in \cite{BKWLetter}, using a toy example of the Bessel--Clifford function $\calK_\alpha(z)$ (for the reader's convenience, we reproduce the corresponding reasoning in the Appendix~\ref{BesselCliffordAppendix}) and for the simplest operator functions $\hat F^{-\mu}$ and $\exp(-\tau\hat F^\nu)$, we show that after regularization using analytic continuation, the expansion \eqref{general_kernel_series_rep} is not meaningless at all, but allows for a transparent physical interpretation. It turns out that since the DeWitt expansion \eqref{HeatKernelExpansion} describes only the UV behavior of the heat kernel $\hat K_F(\tau | x,x')$ as $\tau\to0$, but not its IR behavior as $\tau\to\infty$, its term-by-term integration allows us to obtain not the total expansion for the kernel $\hat K[f(F) | x,x']$, but only half of its terms coming from the UV region. If we wanted to obtain the other, IR half of the expansion for the kernel $\hat K[f(F) | x,x']$, we would need to integrate term-by-term the asymptotic expansion of the heat kernel $\hat K_F(\tau | x,x')$ as $\tau\to\infty$. Such an IR expansion of the heat kernel can be obtained by resumming the DeWitt series \cite{CPTII, Barvinsky2002, Barvinsky03}. However, a detailed consideration of these IR terms is a matter for future research and is beyond the scope of this paper, in which we concentrate entirely on the study of UV terms for operator functions of a more complex form than those considered in \cite{BKWLetter}.

Now let us discuss in more detail the issue of regularization of IR divergent integrals in \eqref{frakLtransformEq2}. We use dimensional regularization throughout, in which the spacetime dimension $d$ can formally take not only positive integer values, but also arbitrary complex values. In addition, we apply regularization via analytic continuation with respect to the parameter $\alpha$ in \eqref{frakLtransformEq2}: if the integrals converge in some region of parameter values, then analytic continuation of the resulting functions beyond this region yields a regularized value of the divergent integral.\footnote{Note that although this approach may initially seem weird, it can be rigorously justified within the Mellin--Barnes framework we use, as a deformation from straight to curved integration contours.} These methods, of course, do not lead to the complete disappearance of IR divergences, but only to their localization in the form of certain poles with respect to the parameter $\alpha$.

The following circumstance may cause some concern: since the expansion \eqref{general_kernel_series_rep} includes terms obtained by integrating increasingly higher powers of the proper time $\tau$, their convergence properties at the IR limit $\tau=\infty$ also deteriorate. Trouble comes in batches: if physical IR divergences arise, they arise simultaneously in an infinite number of terms in the expansion \eqref{general_kernel_series_rep}. This reinforces doubts about the applicability of the method used and prompts a desire to verify the obtained results using a back up regularization method that would completely eliminate all IR divergences and make all integrals well-defined.

The simplest way of such IR regularization is to introduce a mass term $m^2$ into the operator $\hat F(\nabla)\mapsto \hat F(\nabla)+m^2\hat 1$. Then the associated decreasing factor $\exp(-\tau m^2)$ will immediately ensure good convergence of all integrals on the IR limit $\tau=\infty$. More specifically, this leads to the following form of the DeWitt expansion \eqref{HeatKernelExpansion} (in the formulas below, we omit the dependence on the points $x$ and $x'$ for brevity)
\begin{align} 
\hat K_{F+m^2}(\tau) &= \sum\limits_{k=0}^\infty W_{\tfrac{d}{2} - k}(\tau,\sigma, m^2) \cdot \hat a_k[F], \label{ModifiedHKexpansion} \\
W_\alpha(\tau,\sigma, m^2) &= \frac{\tau^{-\alpha}}{(4\pi)^{d/2}} \exp\left(-\frac{\sigma}{2\tau} - m^2 \tau\right). \label{IRregKernel}
\end{align}
Then, term-by-term application of the integral transform $\frakL_f$ \eqref{defLf} to the expansion \eqref{ModifiedHKexpansion} leads to the following modification of the expansion \eqref{general_kernel_series_rep}:
\begin{align}
\hat K\!\big[f(F+m^2)\big] &= \sum\limits_{k=0}^\infty \bbW_{\tfrac{d}{2}-k}[f | \sigma, m^2] \cdot \hat a_k[F], \label{ModifiedExpansion} \\
\bbW_\alpha[f | \sigma, m^2] &= \frakL_f W_\alpha(\tau,\sigma, m^2)  \label{BHtransform} \\
&= \int\limits_0^\infty \frac{d\tau f^*(\tau)\,\tau^{-\alpha}}{(4\pi)^{d/2}} \exp\left(\!-\frac{\sigma}{2\tau} \!-\! m^2 \tau\!\right) \nonumber
\end{align}
We will call the new objects $\bbW_\alpha[f | \sigma, m^2]$ \emph{complete massive kernels}. On the one hand, they do not carry any new information compared to the basis kernels $\bbB_\alpha[f | \sigma]$ (because, obviously, $\bbW_\alpha[f(F) | \sigma, m^2] = \bbB_\alpha[f(F+m^2) | \sigma]$\footnote{Note that $F$ is a dummy variable here. In fact, the kernels $\bbB_{\a}$, $\bbM_{\a}$, and $\bbW_{\a}$ do not depend on the operator $\hat F$. We do not write hats over dummy variables (and also in the HaMiDeW coefficients $\hat a_k[F]$).}), however, at the same time, they are well defined in all cases and, in particular, do not have any poles in the parameter $\alpha$ associated with IR divergences.

After we calculate the complete kernels $\bbW_\alpha[f | \sigma, m^2]$, it is important to learn how to go to the coincidence limit $\sigma\to0$ in them (since it is this limit that determines the expressions for the functional trace and the effective action \eqref{effective_action}-\eqref{FuncTrDef}). In this regard, it is convenient to introduce a third set of objects
\begin{align}
\bbM_\alpha\!\big[f \big| m^2\big] &= \frakL_f M_\alpha(\tau,m^2), \label{MassiveKernelsDef} \\
M_\alpha(\tau,m^2) &= \frac{\tau^{-\alpha}}{(4\pi)^{d/2}}\,\exp(-\tau m^2). \label{ComplInitEq}
\end{align}
which we will call \emph{complementary kernels} and which control the coincidence limit. In their properties, they are indeed in many ways complementary to the basis kernels $\bbB_\alpha[f | \sigma]$ \eqref{frakLtransformEq2}. For example, the integrals in \eqref{MassiveKernelsDef} can also diverge, but not at the IR limit $\tau=\infty$, but at the opposite UV limit $\tau=0$, and also require a regularization procedure using analytic continuation for their definition.

The three types of objects introduced---basis $\bbB_\alpha\![f | \sigma]$, complementary $\bbM_\alpha\!\big[f \big| m^2\big]$, and complete $\bbW_\alpha\!\big[f \big| \sigma, m^2\big]$ kernels---are the centerpiece of the present paper. They connect our two approaches—analytic continuation and mass-term regularization—in a unified framework. We further demonstrate this with examples, computing these objects for operator functions of three types: $\exp(-\tau\hat F^\nu)/\hat F^\mu$, $1/(\hat F^\mu + \lambda)$, and, finally, $\exp(-\tau\hat F^\nu)/(\hat F^\mu + \lambda)$. A remarkable feature of our approach is that in each of these cases (as well as for more complex functions), the kernels can be universally expressed in terms of multiple Mellin-Barnes (MB) integrals.

$N$-fold MB integrals \cite{Zhdanov1998, Ananthanarayan2021} are remarkable special functions that have attracted increasing attention from physicists in recent years because all Feynman integrals can be expressed in terms of them (the so-called Lee--Pomeransky representation \cite{Lee2013, Dubovyk2022}), and they also arise in many other problems. They can be defined as functions of $N$ complex variables of the following form
\begin{equation}\label{NfoldMBint}
H(\bm{z})=\int\limits_{C} \frac{d^Ns}{(2\pi i)^N}\, h(\bm{s})\, \bm{z}^{-\bm s},
\end{equation}
where $\bm{z}, \bm{s} \in\bbC^N$, $\bm{z}^{-\bm s} = \prod\nolimits_{a=1}^N z_a^{-s_a}$, $h(\bm{s})$ is some ratio of gamma functions with arguments linear in $\bm{s}$ and $C$ is a certain $N$-cycle. In our case, $N$ is one less than the total number of independent dimensional parameters of the problem, and $z_a$ are some dimensionless combinations of these parameters.

The advantage of the MB representation is that it allows one to build the asymptotic expansion of our kernels as a series of residue contributions originating from the poles of their Mellin images $h(\bm{s})$. We present these expansions for some simple cases of single argument functions in Sec.~\ref{MainSection}. For functions of multiple arguments, the series representations become more complicated, especially for the so-called resonant cases with merging poles of gamma functions, and their derivation is relegated to the accompanying paper~\cite{BKW25b}.

Obtained expansions of the operator function kernels show full consistency of the results achieved within analytic and mass parameter regularization. As is usually expected from dimensional regularization, which nullifies power divergences and reveals logarithmic divergences in the form of poles in spacetime dimensionality, the analytic regularization yields pole contributions exactly corresponding to the logarithmic in mass terms of the massive regularization. But it totally annihilates the sequence of power singularities in inverse powers of the mass parameter $1/(m^2)^n\to\infty$ arising in the massless limit $m^2\to 0$. More interesting is the confirmation of the observation \cite{Barvinsky25} that the class of functions $f(\lambda)$ non-singular at $\lambda=0$ does not lead to IR divergences for kernels of operator function $f(\hat F)$. This observation was done for the heat kernel of rather generic nonminimal operators of elliptic nature forming a well-defined bounded operator. Their building blocks of the form (\ref{Operator}) always enter the final answer in such combinations that their immanent IR divergences completely cancel out. This was proven in \cite{Barvinsky25} by the special procedure of IR subtraction for hybrid operators (\ref{Operator}) whose finiteness is now confirmed within both analytical and mass-type regularizations.

The structure of the present paper is as follows: in section ~\ref{IntegralTransformsSec} we discuss in more detail operator functions, their kernels, and the integral transforms that relate them. After that, we consider the simplest applications of the off-diagonal functoriality in Sec.~\ref{MainSection}, and then in Sec.~\ref{ResultsSection} we consider several more complicated examples leading to 2- and 3-fold MB integrals. In these sections, we use dimensional regularization throughout, due to which the expressions we obtain will pertain to the so-called non-resonant case. Removing dimensional regularization, i.e., passing to the integer physical dimension $d$, corresponds to the transition to the resonant case: the origin of either contour pinches and associated divergences, or logarithmic terms in the expansion series. These cases will be considered in detail in section~\ref{ResonantCaseSec}. In appendices, we briefly present information on the Bessel--Clifford functions $\calK_\alpha(z)$ and provide a summary of representations for the obtained kernels along with their various limits.

\section{Operator functions, their kernels, and integral transforms} \label{IntegralTransformsSec}

In this section, we take a closer look at operator functions, their integral kernels, and the integral transforms that relate various operator functions and kernels. In the Introduction, we have already mentioned perhaps the most important operator function---the operator exponential, defined as the sum of the corresponding Taylor series
\begin{equation} \label{OpExpDef}
\exp(-\tau\hat F) = \sum\limits_{n=0}^\infty \frac{1}{n!} (-\tau\hat F)^n.
\end{equation}

Two other very important operator functions are the resolvent $(\hat F - \lambda)^{-1}$ and the complex power $\hat F^{-\mu}$. The meaning of the resolvent is intuitive: the inverse operator is defined in the usual way $\hat F \hat F^{-1} = \hat F^{-1}\hat F = \hat 1$, and the resolvent is well defined for any $\lambda$ not belonging to the spectrum of $\hat F(\nabla)$. At the same time, meaning of $\hat F^{-\mu}$ for non-integer, and especially complex, values of $\mu$ may not be entirely obvious. It is standard \cite{Seeley} to define the complex power by using, what is known in the QFT context as Schwinger representation,
\begin{equation} \label{CompPowDef}
\hat F^{-\mu} = \frac{1}{\Gamma(\mu)} \int\limits_0^\infty d\tau\; \tau^{\mu-1}\, e^{-\tau\hat F},
\end{equation}
where $\mu\ne 0, -1, -2, \ldots$. This definition is indeed natural because, using the beta function it is easy to verify that \eqref{CompPowDef} satisfies $\hat F^a\hat F^b = \hat F^{a+b}$, and by repeatedly integrating by parts, that it coincides with the power for positive integer $\mu$.

It is also easy to write out the inverse relation to \eqref{CompPowDef}:
\begin{equation} \label{CompPowInw}
e^{-\tau\hat F} = \int\limits_{w-i\infty}^{w+i\infty} \frac{d\mu}{2\pi i}\, \tau^{-\mu}\, \Gamma(\mu) \,\hat F^{-\mu},
\end{equation}
where $w>0$. It can be interpreted as follows: closing the vertical integration contour on the left allows us to reduce the integral to the sum of residues at the poles of the gamma function, which again reproduces the Taylor series \eqref{OpExpDef}.

It is just as easy to find an integral transform that allows us to obtain the resolvent from the complex power:
\begin{equation} \label{PowerToRes}
\frac{1}{\hat F + \lambda} = \int\limits_C \frac{ds}{2\pi i}\, \Gamma(1-s) \Gamma(s)\; \lambda^{-s} \hat F^{s-1},
\end{equation}
where the contour $C$ separates the poles of the two gamma functions. Indeed, by closing it on the left again and reducing the integral to the sum of residues at the poles of $\Gamma(s)$, we simply reproduce the well-known Neumann operator series:
\begin{equation}
\frac{1}{\hat F + \lambda} = \frac{\hat F^{-1}}{1 + \lambda \hat F^{-1}} = \frac{1}{\hat F} \sum\limits_{k=0}^\infty \left(-\frac{\lambda}{\hat F}\right)^k.
\end{equation}

In fact, we can go further and express any operator function $f(\hat F)$ as a linear integral transform of each of the three operator functions described above. First of all, as is well-known from spectral theory, every operator function can be represented as the following integral of the resolvent:
\begin{equation} \label{ResolventInt}
f(\hat F) = \int\limits_C \frac{d\lambda}{2\pi i} \frac{f(\lambda)}{\lambda - \hat F},
\end{equation}
where the contour $C$ goes around the spectrum of the operator $\hat F$, i.e. around the positive real semi-axis for a non-negatively definite Hermitian operator. But we could equally well express $f(\hat F)$ in terms of the operator exponential. To do this, it is sufficient to express the resolvent in terms of it
\begin{equation} \label{KernelToResolvent}
\frac{1}{\hat F - \lambda} = \int\limits_0^\infty d\tau\; e^{-\tau(\hat F - \lambda)},
\end{equation}
substitute this relation into \eqref{ResolventInt} and change the order of integration. We obtain the representation \eqref{frakLtransformEq}:
\begin{align}
f(\hat F) &= -\int\limits_C \frac{d\lambda}{2\pi i} \int\limits_0^\infty d\tau\; e^{-\tau(\hat F - \lambda)} f(\lambda) \nonumber \\
&= \int\limits_0^\infty d\tau\; e^{-\tau\hat F} f^*(\tau), \label{GenIntTransform}
\end{align}
where $f^*(\tau)$ is the inverse Laplace transform (the sign change in the second line of~\eqref{GenIntTransform} is due to differing contour orientation conventions in the spectral representation \eqref{ResolventInt} and inverse Laplace transform~\eqref{InvLaplaceTransform}). It is important to note that the kernels of integral operator transforms and their inverse transforms should be understood in a generalized sense, i.e. not as functions, but distributions. We will encounter an illustration of such a situation below, see \eqref{MellinConvolution}--\eqref{NuKernel}.

It is often convenient to reformulate statements about operator functions in the synonymous language of integral kernels \eqref{pdo_kernel}. Besides the heat kernel, i.e. the kernel of the operator exponential $\hat K_F(\tau | x,x')$ \eqref{HeatKernelDef}, another important kernel is the Green function, i.e. the kernel of the inverse operator
\begin{equation}
\hat G_F(x,x') = \hat K[F^{-1} | x,x'].
\end{equation}
The kernels of the resolvent and the complex power of the operator are simply the Green functions $\hat G_{F-\lambda}(x,x')$ and $\hat G_{F^\mu}(x,x')$.

To rewrite the relations \eqref{CompPowDef} and \eqref{CompPowInw} in the language of integral kernels, we should just act by them upon the delta function $\delta(x,x')$. We obtain:
\begin{align}
\hat G_{F^\mu}(x,x') &= \frac{1}{\Gamma(\mu)} \int\limits_0^\infty d\tau\; \tau^{\mu-1} \hat K_F(\tau | x,x'), \label{CompPowRep} \\
\hat K_F(\tau | x,x') &= \int\limits_{w-i\infty}^{w+i\infty} \frac{d\mu}{2\pi i}\, \tau^{-\mu}\,\Gamma(\mu)\,\hat G_{F^\mu}(x,x'), \label{CompPowRep2}
\end{align}
where $\mu\neq 0,-1,-2,\dots$ and $w>0$. This means that the heat kernel and the complex power kernel are related to each other by the direct and inverse Mellin transforms in the variables $\tau$ and $\mu$ respectively.

Recall that the direct $\frakM$ and inverse $\frakM^{-1}$ Mellin transforms with respect to complex variables $z$ and $s$ are defined as
\begin{align}
&h(s) = (\frakM H)(s) = \int\limits_0^\infty dz\,z^{s-1} H(z),\label{DirectMellinDef0}\\
&H(z) = (\frakM^{-1}h)(z) = \int\limits_{w-i\infty}^{w+i\infty} \frac{ds}{2\pi i}\, z^{-s} h(s), \label{InverseMellinDef0}
\end{align}
where the straight vertical integration contour lies in the fundamental strip of the function $H(z)$. We discuss Mellin transforms and their properties in detail in the accompanying paper \cite{BKW25b}. From now on, we will denote the original function with an uppercase letter and its Mellin image with a lowercase letter.

Similarly, rewriting the relation \eqref{ResolventInt} for the operator exponential and its inverse \eqref{KernelToResolvent} in the language of kernels, we obtain ($w>0$):
\begin{align}
\hat G_{F+\lambda}(x,x') &= \int\limits_0^\infty d\tau\; e^{-\lambda\tau} \hat K_F(\tau | x,x'), \label{LaplaceRep} \\
\hat K_F(\tau | x,x') &= \int\limits_{w-i\infty}^{w+i\infty} \frac{d\lambda}{2\pi i}\, e^{\lambda\tau} \hat G_{F+\lambda}(x,x'). \label{LaplaceRep2}
\end{align}
This means that the heat kernel and the resolvent kernel are related to each other by the direct and inverse Laplace transforms ({denoted by $\frakL$}) in the variables $\tau$ and $\lambda$.

Finally, for the transformation \eqref{PowerToRes} and the relation \eqref{ResolventInt} for the operator complex power, we have:
\begin{align}
\hat G_{F+\lambda}(x,x') &=  \int\limits_C \frac{ds}{2\pi i} \lambda^{-s}\Gamma(1-s)\Gamma(s)\, \hat G_{F^{1-s}}(x,x'), \label{NonameTrans1} \\
\hat G_{F^\mu}(x,x') &= \int\limits_{w-i\infty}^{w+i\infty} \frac{d\lambda}{2\pi i} (-\lambda)^{-\mu} \hat G_{F+\lambda}(x,x'), \label{NonameTrans2}
\end{align}
where the contour $C$ separates the poles of the two gamma functions and $w>0$. As far as we know, these transforms do not bear a special name.

In general, the relations between all three kernels given by the transforms \eqref{CompPowRep}-\eqref{NonameTrans2} can be summarized in the following simple diagram:
\begin{equation} \label{fund_relations_kernels}
\xymatrix{ & \hat K_F(\tau| x,x') \ar@{<->}[dl]_\frakM \ar@{<->}[dr]^\frakL & \\
\hat G_{F^\mu}(x,x') \ar@{<->}[rr] && G_{F+\lambda}(x,x'),
}\end{equation}
where $\frakM$ and $\frakL$ denote the Mellin and Laplace transforms, respectively, and the unlabeled lower arrow denotes the transforms \eqref{NonameTrans1}-\eqref{NonameTrans2}.

\section{The simplest examples} \label{MainSection}

In this section, we look at some simple considerations that we first used in \cite{Wach3}. We now view them as the simplest applications of the much more general idea of off-diagonal functoriality that we outlined in the Introduction, and as a basis for considering more complex cases in Section~\ref{ResultsSection}.

First, a technical remark is in order: by substituting an explicit expression for the integral transform $\frakL_f$ \eqref{frakLtransformEq} into the definition of the basis kernel $\bbB_\alpha[f | \sigma]$ \eqref{frakLtransformEq2}, as a convolution with the inverse Laplace transform $f^*(\tau)$ \eqref{InvLaplaceTransform}, we obtain the following general integral representation for basis kernels
\begin{align}
&\bbB_\alpha[f | \sigma] = \frakL_f B_\alpha(\tau, \sigma) = \int\limits_0^\infty d\tau\, f^*(\tau) B_{\alpha}(\tau,\sigma) \nonumber \\
&=\int\limits_C \frac{d\lambda}{2\pi i}\, f(\lambda) \int\limits_0^\infty \frac{d\tau\, \tau^{-\alpha}}{(4\pi)^{d/2}} \exp\Big(\tau\lambda - \frac{\sigma}{2\tau}\Big) \nonumber \\
&= \int\limits_C \frac{d\lambda}{2\pi i}\, f(\lambda) \frac{2(-\lambda)^{\alpha-1}}{(4\pi)^{d/2}} \calK_{\alpha-1}\!\Big(-\frac{\lambda\sigma}{2}\Big), \label{GenBasis1}
\end{align}
where $\calK_\alpha(z)$ is the Bessel--Clifford function of second kind \eqref{BC2Def}. In other words, we can immediately write the general answer for an arbitrary basis kernel $\bbB_\alpha[f | \sigma]$ as a contour integral over the spectral parameter $\lambda$ of the product of the function $f(\lambda)$ and, essentially, $\calK_\alpha(-\lambda\sigma/2)$. Of course, one can construct similar representations for massive $\bbM_\alpha[f | m^2]$ \eqref{MassiveKernelsDef} and complete $\bbW_\alpha[f | \sigma, m^2]$ \eqref{BHtransform} kernels as well.

The problem is that general answers such as \eqref{GenBasis1} are not very convenient in practice. Instead, as we will see more than once in what follows, the power of the integral transform technique developed is due to the remarkable properties of MB integrals. Accordingly, we should strive to represent all the answers we obtain in the MB integral form. Representations of the form \eqref{GenBasis1} not only do not have this form themselves, but can also make it nontrivial to arrive at such a representation.

Therefore, we take another, more convenient approach: first, break down the integral transform $\frakL_f$ mapping the operator exponential $\exp({-\tau\hat F})$ into the desired operator function $f(\hat F)$ into a number of elementary steps. Then, sequentially apply these simpler transforms to the corresponding kernels, each time expressing the answer as a MB integral. This is precisely the method we will use in calculations that follow.

\subsection{Basis kernels for $\hat F^{-\mu}$ and $\exp(-\tau\hat F^\nu)$}

Let us start with considering the simplest example: a basis kernel for a complex power $\hat F^{-\mu}$. Our general representation \eqref{general_kernel_series_rep}-\eqref{frakLtransformEq2} in this case reduces to substituting the DeWitt expansion \eqref{HeatKernelExpansion} into the Mellin transform \eqref{CompPowRep} and takes the following form:
\begin{equation} \label{HadamardExpansion}
\hat G_{F^\mu}(x,x') = \sum\limits_{k=0}^\infty \bbB_{\frac{d}{2} - k}\!\big[F^{-\mu} \big| \sigma\big] \cdot \hat a_k[F | x,x'],
\end{equation}
where
\begin{align}
\bbB_\alpha\!\big[F^{-\mu} \big| \sigma\big] &= \frac{1}{\Gamma(\mu)} \int\limits_0^\infty d\tau\; \tau^{\mu-1} B_\alpha(\tau,\sigma) \nonumber \\
&= \frac{1}{(4\pi)^{d/2}\Gamma(\mu)} \int\limits_0^\infty d\tau\; \tau^{\mu-\alpha-1} e^{-\sigma/2\tau} \nonumber\\
&= \frac{\left(\sigma/2\right)^{\mu-\alpha}}{(4\pi)^{d/2}\Gamma(\mu)} \int\limits_0^\infty dz\; z^{\alpha-\mu-1}\; e^{-z} \nonumber \\
&= \frac{\Gamma\left(\alpha-\mu\right)}{(4\pi)^{d/2}\Gamma(\mu)} \left(\frac{\sigma}{2}\right)^{\mu-\alpha}. \label{CompPowFunctions}
\end{align}
Here, when passing from the second to the third line, we made the substitution $z = \sigma/2\tau$.

In this example we first encounter IR divergences, the problem mentioned in the Introduction. Indeed, due to the presence of the exponential factor $e^{-\sigma/2\tau}$, the proper time integral~(\ref{CompPowFunctions}) always converges at the UV limit $\tau=0$, but diverges at the IR limit $\tau = \infty$ for
\begin{equation} \label{ConvergenceRegion}
\Re(\alpha-\mu) < 0.
\end{equation}
Since we will encounter this condition repeatedly in what follows, we assign a name to the region of parameters defined via~(\ref{ConvergenceRegion}). Borrowing terminology from the renormalization theory, it is appropriate to call this parameter region \emph{irrelevant}, and to call the complementary region $\Re(\alpha-\mu)>0$ \emph{relevant}. And since the parameter $\alpha = \tfrac{d}{2} - m$ takes on infinitely decreasing values, the expansion~(\ref{HadamardExpansion}) contains only a finite number of ``relevant'' terms and an infinite number of ``irrelevant'' divergent terms. Therefore, the divergence of integrals seems to pose a significant problem.

We will understand the expression \eqref{CompPowFunctions} in the sense of the analytic continuation rule  we formulated in Introduction, i.e., in other words, we simply postulate that the basis kernel $\bbB_\alpha\!\big[F^{-\mu} \big| \sigma\big]$ are given by the formula \eqref{CompPowFunctions} even in cases where the parameter $\alpha$ lies in the irrelevant region. Note that this assumption in itself does not eliminate IR divergences as such, but only allows us to work with them in a controlled manner. Indeed, even after such regularization, the expression \eqref{CompPowFunctions} retains divergences: it becomes infinite when the parameter combination $\alpha-\mu$ takes on non-positive integer values. In particular, for even $d$ and integer $\mu$ or for odd $d$ and half-integer $\mu$ in an infinite number of terms of \eqref{HadamardExpansion}, the factor $\Gamma(\alpha-\mu)$ has a pole. These divergences are not an artifact of our method, but are physical IR divergences.

Does this strategy yield any meaningful answers? Yes, and to see this, it is enough to go just one step further and consider the operator's power exponential $\exp({-\tau\hat F^\nu})$. Its general representation \eqref{general_kernel_series_rep}-\eqref{frakLtransformEq2} (obtained by substituting the formulas \eqref{HadamardExpansion}-\eqref{CompPowFunctions} into the inverse Mellin transform \eqref{CompPowRep2}) takes the form:
\begin{equation} \label{FnuHeatKernel}
\hat K_{F^\nu}(\tau | x,x') = \sum\limits_{k=0}^\infty \bbB_{\frac{d}{2} - k}\!\big[e^{-\tau F^\nu} \big| \sigma\big] \cdot \hat a_k[F | x,x'],
\end{equation}
where
\begin{align}
&\bbB_\alpha\!\big[e^{-\tau F^\nu} \big| \sigma\big] = \int\limits_C \frac{d\rho}{2\pi i}\, \tau^{-\rho}\,\Gamma(\rho)\;\bbB_\alpha\!\left[F^{-\rho\nu} | \sigma\right] \nonumber \\
&= \int\limits_C \frac{d\rho}{2\pi i}\, \frac{\Gamma(\rho)\Gamma(\alpha-\rho\nu)}{(4\pi)^{d/2}\, \Gamma(\rho\nu)}\, \tau^{-\rho}\left(\frac{\sigma}{2}\right)^{\rho\nu-\alpha} \nonumber \\
&= \frac{\tau^{-\frac{\alpha}{\nu}}}{(4\pi)^{d/2}} \int\limits_{C'} \frac{ds}{2\pi i}\, \frac{\Gamma(s)\Gamma\left(\frac{\alpha-s}{\nu}\right)}{\nu\Gamma(\alpha-s)} \left(\frac{\sigma}{2\tau^{1/\nu}}\right)^{-s} \nonumber \\
&= \frac{\tau^{-\frac{\alpha}{\nu}}}{(4\pi)^{d/2}}\, \calE_{\nu,\alpha}\!\left(-\frac{\sigma}{2\tau^{1/\nu}}\right). \label{PowHeatKernel}
\end{align}
Here, when passing from the second to the third line, we made the substitution $s = \alpha - \rho\nu$, and then reduced the result to \emph{generalized exponential functions (GEFs)} $\calE_{\nu,\alpha}(z)$ which were introduced in \cite{Wach2}. These special functions can be defined in terms of the MB integral:
\begin{align}
\calE_{\nu,\alpha}(z) &= \int\limits_C \frac{ds}{2\pi i}\,(-z)^{-s}\, \varepsilon_{\nu,\alpha}(s) \nonumber \\
&= \frac{1}{\nu} \sum\limits_{m=0}^\infty \frac{\Gamma\left(\frac{\alpha+m}{\nu}\right)}{\Gamma(\alpha+m)} \frac{z^m}{m!}, \label{InvMellinCalE} \\
\varepsilon_{\nu,\alpha}(s) &= \int\limits_0^\infty dz\, z^{s-1}\,\calE_{\nu, \alpha}(-z) = \frac{\Gamma(s)\Gamma\left(\frac{\alpha-s}{\nu}\right)}{\nu\Gamma(\alpha-s)}. \label{MellinCalE}
\end{align}
Their properties were studied in detail in our paper~\cite{Wach2}, to which we refer the interested reader. At the moment, it is important for us that the expressions \eqref{PowHeatKernel}-\eqref{InvMellinCalE} are well defined and no longer contain any divergences for positive integer $\nu$.

In particular, if we define the Seeley--Gilkey coefficients of the operator $\hat H(\nabla)$ of order $2\nu$ as the coefficients of its heat kernel diagonal ($x=x'$) expansion in powers of the proper time,
\begin{equation}
\hat K_H(\tau | x, x) = \tau^{-d/2\nu} \sum\limits_{n=0}^\infty \tau^{n/\nu}\, \hat E_{2n}[H| x],
\end{equation}
then, taking the coincidence limit $\sigma\to0$ \eqref{FnuHeatKernel} and using $\calE_{\nu,\alpha}(0) = \Gamma(\alpha/\nu)/\nu\Gamma(\alpha)$, we restore the well-known Fegan--Gilkey formula \cite{GilkeyFegan}
\begin{equation} \label{FeganGilkey}
\hat E_n[F^\nu | x] = \frac{\Gamma\left(\frac{d - n}{2\nu}\right)}{\nu\Gamma\left(\frac{d - n}{2}\right)}\, \hat E_n[F | x].
\end{equation}

Thus, our off-diagonal method reproduces the correct diagonal result \eqref{FeganGilkey}. Moreover, the GEFs $\calE_{\n,\a}(-z)$ turn out to be extremely important: they appear in the off-diagonal heat kernel expansion for a general minimal operator of higher order, which is a generalization of \eqref{FnuHeatKernel}.\footnote{It also contains an infinite number of additional terms with negative powers of the proper time $\tau$, the coefficients of which, however, vanish in the coincidence limit.} We had previously verified this result by two completely independent methods: using the ``generalized Fourier transform'' in curved spacetime in \cite{Wach3} and perturbation theory in \cite{BKW2024}.

Finally, one more important remark should be made. One can use the direct \eqref{CompPowDef} and inverse \eqref{CompPowInw} Mellin transforms to express $\exp({-\tau\hat F{}^\nu})$ straightforwardly as an integral transform of the original $\exp({-\tau\hat F})$. Indeed,
\begin{align}
e^{-\tau\hat F^\nu}& = \frac{1}{2\pi i} \int\limits_{w-i\infty}^{w+i\infty} d\mu\; \Gamma(\mu)\, \tau^{-\mu} \hat F^{-\mu\nu} \nonumber \\
&= \frac{1}{2\pi i} \int\limits_{w-i\infty}^{w+i\infty} d\mu\; \frac{\Gamma(\mu)}{\Gamma(\mu\nu)}\, \tau^{-\mu} \int\limits_0^\infty dt\; t^{\mu\nu-1}  e^{-t\hat F} \nonumber \\
&= \int\limits_0^\infty \frac{dt}{t} e^{-t\hat F} \left[\frac{1}{2\pi i} \int\limits_{w-i\infty}^{w+i\infty} d\mu\; \frac{\Gamma(\mu)}{\Gamma(\mu\nu)}\, \Big(\frac{\tau^{1/\nu}}{t}\Big)^{-\mu\nu}\right] \nonumber \\
&= \int\limits_0^\infty \frac{dt}{t}\; \phi_\nu\!\Big(\frac{\tau^{1/\nu}}{t}\Big)\, e^{-t\hat F}, \label{MellinConvolution}
\end{align}
where the kernel $\phi_\nu(\tau^{1/\nu}/t)$ of this transform is given by the function
\begin{equation} \label{NuKernel}
\phi_\nu(z) = \int\limits_{w'-i\infty}^{w'+i\infty} \frac{ds}{2\pi i}\; \frac{\Gamma(s/\nu)}{{\nu}\Gamma(s)}\, z^{-s}.
\end{equation}
This representation deserves two comments.

First, the integral \eqref{MellinConvolution} can be interpreted as a so-called Mellin convolution
\begin{equation}
(f * g)(z)=\int\limits_0^{\infty}\frac{dt}{t}\,f\!\left(\frac{z}{t}\right)\, g(t).
\end{equation}
This operation has a remarkable property \cite{Marichev1983}: the Mellin image of the Mellin convolution of two functions is equal to the product of their Mellin images
\begin{equation} \label{ConvolutionProperty}
\frakM(f * g) = (\frakM f) \cdot (\frakM g).
\end{equation}
On the other hand, one can verify (see the accompanying paper \cite{BKW25b}) that for $\nu\ge1$ the integral \eqref{NuKernel} diverges for all values of $z$. Therefore, the kernel $\phi_\nu(\tau^{1/\nu}/t)$ should be understood not as an ordinary function, but in a generalized sense as a distribution (similar to how the delta function can be represented as a Fourier integral). Its Mellin convolutions \eqref{MellinConvolution} with test functions can be defined purely formally: as the multiplication of the Mellin image of the original function by the factor $\frakM\phi_\nu = \Gamma(s/\nu)/\nu\Gamma(s)$, in accordance with the property \eqref{ConvolutionProperty}.

\subsection{Complementary kernels for $\hat F^{-\mu}$ and $\exp({-\tau\hat F^\nu})$}

Now we obtain expressions complementary to the formulas \eqref{CompPowFunctions} and \eqref{PowHeatKernel}. Substituting the expression $M_{\alpha}(\tau,m^2)$ \eqref{ComplInitEq} into \eqref{CompPowFunctions} instead of $B_{\alpha}(\sigma,\tau)$ \eqref{InitialKernel}, we obtain:
\begin{align}
\bbM_\alpha\!\big[F^{-\mu} \big| m^2\big] &= \frac{1}{\Gamma(\mu)} \int\limits_0^\infty d\tau\; \tau^{\mu-1} M_\alpha(\tau,m^2) \nonumber \\
&= \frac{1}{(4\pi)^{d/2}\Gamma(\mu)} \int\limits_0^\infty d\tau\; \tau^{\mu-\alpha-1} e^{-m^2\tau} \nonumber \\
&= \frac{\Gamma\left(\mu-\alpha\right)}{(4\pi)^{d/2}\Gamma(\mu)} m^{2(\alpha-\mu)}. \label{CompPowFunctions2}
\end{align}
Here, the convergence properties is directly opposite to those we encountered when analyzing expression~\eqref{CompPowFunctions}: due to the presence of the exponential factor $\exp({-m^2\tau})$, the proper time integral always converges at its IR limit $\tau=\infty$, but diverges at the UV limit $\tau = 0$ in the relevant region $\Re(\alpha - \mu) > 0$, and the integral must be analytically continued into this region. Accordingly, the convergence and divergence regions for the complementary kernel switch places.

Similarly, replacing the basis kernels $\bbB_\alpha$ \eqref{CompPowFunctions} in the formula \eqref{PowHeatKernel} with complementary kernels $\bbM_\alpha$ \eqref{CompPowFunctions2}, we obtain
\begin{align}
&\bbM_\alpha\!\big[e^{-\tau F^\nu} \big| m^2\big] = \int\limits_C \frac{d\rho}{2\pi i}\, \tau^{-\rho}\,\Gamma(\rho)\;\bbM_\alpha\!\big[ F^{-\rho\nu} \big| m^2\big] \nonumber \\
&= \frac{m^{2\alpha}}{(4\pi)^{d/2}} \int\limits_C \frac{d\rho}{2\pi i}\, \frac{\Gamma(\rho)\Gamma(\nu\rho-\alpha)}{\Gamma(\nu\rho)}\, \left(m^{2\nu}\tau\right)^{-\rho} \nonumber \\
&= \frac{\tau^{-\frac{\alpha}{\nu}}}{(4\pi)^{d/2}}\, \tilde\calE_{\nu,\alpha}\!\left(m^2\tau^{1/\nu}\right), \label{PowHeatKernel2}
\end{align}
where the new special function $\tilde\calE_{\nu,\alpha}(z)$ is defined using 1-fold MB integral
\begin{equation} \label{tildeEDef}
\tilde\calE_{\nu,\alpha}(z) = \int\limits_C \frac{ds}{2\pi i} \frac{\Gamma(s)\,\Gamma\left(\frac{s+\alpha}{\nu}\right)}{\nu\Gamma(s+\alpha)}\, z^{-s}.
\end{equation}
This function is somewhat similar to the GEF $\calE_{\nu,\alpha}(z)$ \eqref{InvMellinCalE}-\eqref{MellinCalE}, but differs from it in the signs of the variable $s$ inside the arguments of two gamma factors.

\subsection{Complete massive kernels for $\hat F^{-\mu}$}

Substituting the massive kernels \eqref{IRregKernel} into the Mellin transform \eqref{CompPowDef}, we get
\begin{align}
&\bbW_\alpha\!\big[F^{-\mu} \big| \sigma, m^2\big]= \frac{1}{\Gamma(\mu)} \int\limits_0^\infty d\tau\; \tau^{\mu-1}\, W_{\alpha}(\tau,\sigma,m^2) \nonumber \\
&= \frac{m^{2(\alpha-\mu)}}{(4\pi)^{d/2} \Gamma(\mu)} \int\limits_0^\infty dt\; t^{\mu-\alpha-1} \exp\left(-\frac{z}{t} - t\right) \nonumber \\
&= \frac{2m^{2(\alpha-\mu)}}{(4\pi)^{d/2} \Gamma(\mu)}\; \calK_{\alpha-\mu}\!\left(\sigma m^2/2 \right), \label{GreenIRreg}
\end{align}
where, using the substitutions $t = m^2\tau$ and $z = \sigma m^2/2$, we reduced the integral to the Bessel--Clifford function of the 2nd kind \eqref{BC2Def}. The resulting expression is a massive analogue of the formula \eqref{CompPowFunctions}, but unlike it, it is well-defined everywhere, since the integral~\eqref{GreenIRreg} converges well both at the UV limit $\tau=0$ and at the IR limit $\tau=\infty$.

In order to obtain either the coincidence ($\s\to0$) or the massless ($m^2\to0$) limits of the expression~\eqref{GreenIRreg}, we simply substitute the leading term of the asymptotics of $\calK_\alpha(z)$ at $z = \sigma m^2/2 \to0$ \eqref{BCasymptotic} into it. As a result, the answers will differ depending on the sign of $\Re(\a-\m)$.

In the relevant region $\Re(\a-\m)>0$ one has:
\begin{equation} \label{LimitM2to0Conv}
\bbW_\alpha\!\big[ F^{-\mu} \big| \sigma, m^2\big] \xrightarrow[\sigma m^2\to0]{\mathrm{rel}} \bbB_\alpha\!\big[ F^{-\mu} \big| \sigma\big].
\end{equation}
Since the dependence on $m^2$ completely vanishes in this expression, it is exact for the massless limit $m^2\to0$. However, as it is easy to see, the coincidence limit $\sigma\to0$ for this expression is divergent.

On the other hand, in the irrelevant region $\Re(\a-\m)<0$ we obtain:
\begin{equation} \label{DivergenceRegionLimit}
\bbW_\alpha\!\big[ F^{-\mu} \big| \sigma, m^2\big] \xrightarrow[\sigma m^2\to0]{\mathrm{irrel}} \bbM_\alpha\!\big[ F^{-\mu} \big| m^2\big].
\end{equation}
In this expression, the dependence on $\sigma$ completely vanishes, and therefore it is an exact answer for the coincidence limit $\sigma\to0$. However this expression obviously diverges in the massless limit $m^2\to0$. When calculating the quantum effective action in the background field method, an expansion in powers of the background dimension arises just from this expression. It includes all higher heat kernel coefficients divided by powers of the mass parameter $m^2$, which makes terms of this expansion dimensionless. In the framework of covariant perturbation theory \cite{CPTI, CPTII, CPTIIIa, CPTIII}, this infinite series can be resummed into an expansion in powers of curvatures with special nonlocal form factors.

\section{More complex examples} \label{ResultsSection}

In this section, we considered operator functions $f(\hat F)$ of three types: $\exp(-\tau\hat F)/\hat F^\mu$, $1/(\hat F^\mu + \lambda)$, and $\exp(-\tau\hat F)/(\hat F^\mu + \lambda)$. For each of them, we obtained explicit representations in the form of 1-, 2-, and 3-fold MB integrals for the above-introduced basis $\bbB_\alpha\![f(F)| \sigma]$ \eqref{frakLtransformEq2}, complementary $\bbM_\alpha\![f(F)| m^2]$ \eqref{ComplInitEq}, and complete $\bbW_\alpha\![f(F)| \sigma, m^2]$ \eqref{BHtransform} kernels.

There are no fundamental obstacles to applying our method to functions of an arbitrarily more complex form. The reasons why we limit ourselves to the considered functions are, firstly, that we need to stop somewhere, and, secondly, these functions provide a certain initial set of examples of the phenomena that may arise. However, even if we limit ourselves to this small number of examples, their detailed derivation would take up too much space, and it would be easy to get lost in it. Therefore, for the convenience of readers, we have collected all the results we obtained in a compact form in the Appendix~\ref{SummaryAppendix}, and in this section we will limit ourselves to a detailed analysis of several of the most characteristic examples, which sufficiently illustrate the methods we used.

\subsection{Basis kernels for $\exp(-\tau\hat F^\nu)/\hat F^\mu$} \label{Sect4.1}

To calculate any of the kernels, our general method prescribes one to find the integral transform \eqref{frakLtransformEq}, mapping operator function of a simpler form to the required operator function first. Thus, to obtain the operator function $\exp(-\tau\hat F^\nu)/\hat F^\mu$, it suffices to note that, according to \eqref{CompPowDef}, the complex power $\hat F^{-\mu}$ can be constructed not only from $\exp(-\tau\hat F)$, but also from $\exp(-\tau\hat F^\nu)$:
\begin{equation} \label{CompPowDefModified}
\hat F^{-\mu} = \frac{1}{\Gamma(\mu/\nu)} \int\limits_0^\infty dt\; t^{\frac{\mu}{\nu}-1} e^{-t\hat F^\nu}.
\end{equation}
Multiplying this equation by $e^{-\tau\hat F^\nu}$ and shifting the integration variable, we obtain the desired integral transform, which reads
\begin{equation} \label{HybridizingTransform}
\frac{e^{-\tau\hat F^\nu}}{\hat F^\mu} = \frac{1}{\Gamma\left(\mu/\nu\right)} \int\limits_\tau^\infty ds\, (s-\tau)^{\frac{\mu}{\nu}-1}\, e^{-s\hat F^\nu}.
\end{equation}

Now, to obtain the basis kernel $\bbB_\alpha\![\exp(-\tau F^\nu)/F^\mu | \sigma ]$, it is sufficient to simply substitute the previously obtained basis kernel $\bbB_\alpha\![\exp(-\tau F^\nu) | \sigma ]$ \eqref{PowHeatKernel} into the integral transform \eqref{HybridizingTransform}. Thus, we obtain the formula:
\begin{align}
&\bbB_\alpha\!\Big[ \frac{e^{-\tau F^\nu}}{F^\mu} \Big| \sigma\Big] = \frac{1}{\Gamma\left(\frac{\mu}{\nu}\right)} \int\limits_\tau^\infty ds\; (s-\tau)^{\frac{\mu}{\nu}-1} \bbB_\alpha\!\big[ e^{-s F^\nu} \big|  \sigma\big] \nonumber  \\
&= \frac{1}{(4\pi)^{d/2} \Gamma\left(\frac{\mu}{\nu}\right)} \int\limits_\tau^\infty ds\; \frac{(s-\tau)^{\frac{\mu}{\nu}-1}}{s^{\frac{\alpha}{\nu}}} \calE_{\nu,\alpha}\!\left(-\frac{\sigma}{2s^{1/\nu}}\right) \nonumber  \\
&= \frac{\tau^\frac{\mu-\alpha}{\nu}}{(4\pi)^{d/2}\Gamma\left(\frac{\mu}{\nu}\right)} \int\limits_1^\infty dt\; \frac{(t-1)^{\frac{\mu}{\nu}-1}}{t^{\frac{\alpha}{\nu}}} \calE_{\nu,\alpha}\!\left(-\frac{\sigma}{2(t\tau)^{1/\nu}}\right) \nonumber \\
&= \frac{\tau^\frac{\mu-\alpha}{\nu}}{(4\pi)^{d/2}}\; \calE_{\nu,\alpha}^{(\mu)}\!\left(-\frac{\sigma}{2\tau^{1/\nu}}\right). \label{bbBcalE}
\end{align}
Here, when passing from the second to the third line, we made the substitution $t = s/\tau$, and then hid the integral into some new special function
\begin{equation} \label{calK_def}
\calE_{\nu,\alpha}^{(\mu)}(-z) = \frac{1}{\Gamma\left(\frac{\mu}{\nu}\right)} \int\limits_1^\infty dt\, \frac{(t-1)^{\frac{\mu}{\nu}-1}}{t^\frac{\alpha}{\nu}} \calE_{\nu,\alpha}\!\left(-\frac{z}{t^{1/\nu}}\right).
\end{equation}

It is, however, inconvenient to work with the integral representation of the form~\eqref{calK_def}, instead, we want to obtain an expression for $\calE_{\nu,\alpha}^{(\mu)}(-z)$ in the standard form of the MB integral \eqref{calKMellin}. To do this, of course, it is sufficient to simply find the function $\varepsilon_{\nu,\alpha}^{(\mu)}(s)$ which is the Mellin transform of the function $\calE_{\nu,\alpha}^{(\mu)}(-z)$ for which we have:
\begin{align}
&\varepsilon_{\nu,\alpha}^{(\mu)}(s) = \int\limits_0^\infty dz \,z^{s-1} \calE_{\nu,\alpha}^{(\mu)}(-z) \nonumber \\
&= \frac{1}{\Gamma\left(\frac{\mu}{\nu}\right)} \int\limits_1^\infty d\tau\; \frac{(\tau-1)^{\frac{\mu}{\nu}-1}}{\tau^\frac{\alpha}{\nu}} \int\limits_0^\infty dz\; z^{s-1} \calE_{\nu,\alpha}\!\left(-\frac{z}{\tau^{1/\nu}}\right) \nonumber \\
&= \frac{\Gamma(s) \Gamma\left(\frac{\alpha-s}{\nu}\right)}{\nu \Gamma\left(\frac{\mu}{\nu}\right) \Gamma(\alpha-s)} \int\limits_1^\infty d\tau\; (\tau-1)^{\frac{\mu}{\nu}-1} \tau^{\frac{s-\alpha}{\nu}} \nonumber \\
&= \frac{\Gamma(s)\Gamma\left(\frac{\alpha-s-\mu}{\nu}\right)}{\nu\Gamma(\alpha-s)}. \label{AppEq1}
\end{align}
Here we first substituted our original definition of the function $\calE_{\nu,\alpha}^{(\mu)}(-z)$ \eqref{calK_def} and changed the order of integration, then used the already known Mellin transform for the generalized exponential \eqref{MellinCalE} and, finally, reduced the last integral to the beta function.

One can see that the new function $\calE_{\nu,\alpha}^{(\mu)}(z)$ is a modification of the previously introduced function $\calE_{\nu,\alpha}(z)$ \eqref{InvMellinCalE}-\eqref{MellinCalE}, differing from the latter only by the shift of the parameter of one of the gamma functions inside the MB integral by $\mu$. Therefore, as it should be, the limit $\mu\to0$ restores the previously obtained result \eqref{PowHeatKernel}.

The expression~\eqref{AppEq1} contains two rows of poles in $s$, one of which is leftward-running (the one corresponding to the poles of $\Gamma(s)$), and the other is rightward-running (the one corresponding to the two other gamma-functions). We can close the integration contour either to the left or to the right, and, reducing the integral to the sum of residues at the corresponding poles, we obtain the following two expansions:
\begin{align}
\calE_{\nu,\alpha}^{(\mu)}(-z) &= \sum\limits_{n=0}^\infty \frac{\Gamma\left(\frac{n+\alpha-\mu}{\nu}\right)}{\nu\Gamma(n+\alpha)} \frac{(-z)^n}{n!} \label{CalEmu} \\
&= z^{\mu-\alpha} \sum\limits_{n=0}^\infty \frac{\Gamma(\nu n+\alpha-\mu)}{\Gamma(\mu-\nu n)} \frac{(-z^{-\nu})^n}{n!}. \label{CalEmu2}
\end{align}
The first expansion~\eqref{CalEmu} is an everywhere convergent Taylor series at $z=0$, so one has:
\begin{equation} \label{calELim1}
\calE_{\nu,\alpha}^{(\mu)}(0) = \frac{\Gamma\left(\frac{\alpha-\mu}{\nu}\right)}{\nu\Gamma(\alpha)}.
\end{equation}
The second expansion~\eqref{CalEmu2} is an asymptotic series in the vicinity of $z=\infty$. The leading term of the asymptotics is of the form:
\begin{equation} \label{calELim2}
\calE_{\nu,\alpha}^{(\mu)}(-z) \xrightarrow[z\to\infty]{} \frac{\Gamma(\alpha-\mu)}{\Gamma(\mu)}\, z^{\mu-\alpha}.
\end{equation}

Substituting the limit \eqref{calELim2} into the formula \eqref{bbBcalE} and comparing the answer with the expression \eqref{CompPowFunctions}, we obtain the limit \eqref{Lim21Eq}.

\subsection{Complete kernels for $\exp(-\tau\hat F^\nu)/\hat F^\mu$}

To obtain an expression similar to~\eqref{bbBcalE}, but for the complete kernels $\bbW_\alpha$, we must repeat all the same steps, starting not from the basis kernel $\bbB_\alpha\![F^{-\mu}| \sigma]$ \eqref{CompPowFunctions}, but from the complete kernel $\bbW_\alpha\![F^{-\mu}| \sigma, m^2]$ \eqref{GreenIRreg} instead.

First, we obtain the complete kernel $\bbW_\alpha\![\exp(-\tau F^\nu) | \sigma, m^2]$ by substituting \eqref{GreenIRreg} into the integral transformation \eqref{PowHeatKernel}. Substituting then into the resulting expression the representation of the Bessel--Clifford functions $\calK_\nu(z)$ in the form of the MB integral \eqref{BCMeelinBarnes2}-\eqref{BCMeelinBarnes}, we immediately obtain the following expression for the complete massive kernels in the form of a 2-fold MB integral:
\begin{align}
&\bbW_\alpha\!\big[e^{-\tau F^\nu} \big| \sigma, m^2 \big] = \int\limits_C \frac{d\rho}{2\pi i}\, \tau^{-\rho}\,\Gamma(\rho)\;\bbW_\alpha\!\big[ F^{-\rho\nu} \big| \sigma, m^2\big] \nonumber \\
&\qquad\quad= \int\limits_C \frac{d\rho}{2\pi i} \frac{2\Gamma(\rho)\,\tau^{-\rho} m^{2(\alpha-\rho\nu)}}{(4\pi)^{d/2}\Gamma(\rho\nu)}\, \calK_{\alpha-\rho\nu}\!\left(\frac{\sigma m^2}{2}\right) \nonumber \\
&\qquad\quad= \frac{m^{2\alpha}}{(4\pi)^{d/2}}\; H_1\!\left(\frac{\sigma m^2}{2}, m^2\tau^{1/\nu}\right), \label{AppEq2}
\end{align}
where
\begin{align}
H_1\!\left(z_1, z_2\right) &=  \int\limits_C \frac{ds_1 ds_2}{(2\pi i)^2}\, h_1(s_1, s_2)\, z_1^{-s_1} z_2^{-s_2}, \label{2foldMBint} \\
h_1\!(s_1, s_2) &= \int\limits_0^\infty dz_1\, dz_2\, z_1^{s_1-1} z_2^{s_2-1} H_1(z_1, z_2) \nonumber \\
&= \frac{\Gamma(\tfrac{s_2}\nu)}{\nu\Gamma(s_2)} \Gamma(s_1) \Gamma(s_1+s_2-\alpha). \label{h1gamma}
\end{align}

Note that for $\nu=1$ two of the gamma functions in \eqref{h1gamma} cancel each other out. Then the 2-fold integral in \eqref{2foldMBint} decomposes into a product of two 1-fold integrals, each of which yields an exponential, which again brings us to the original expression \eqref{IRregKernel}.

The appearance of $N$-fold MB integrals for operator functions with additional dimensional parameters is a distinctive feature of our approach, which is based on the application of integral transforms. Note that the arguments $z_1, z_2, \ldots$ are dimensionless combinations of the dimensional parameters $m^2$, $\tau$, $\sigma$ etc. So, the integral multiplicity $N$ is one less than the total number of dimensional parameters in the problem. All functions that we will encounter in what follows belong to this very broad class of special functions.

In this paper we do not touch upon the general theory of such special functions leaving its overview for the upcoming paper~\cite{BKW25b}, where this theory is employed to study properties of the functions to follow below. These various properties, including series representations and convergence analysis are simply stated when needed, without derivation, which will be presented in detail in~\cite{BKW25b}.

In what follows, we will also use the following notations and conventions: since $N$-fold MB integrals are a generalization of the well-known Fox $H$-functions~\cite{Marichev1983}, we choose to denote them by the capital letter $H$, and their images under the multiple Mellin transform by the lowercase letter $h$, respectively. In order to distinguish between the various specific functions that we will encounter, we will use different subscripts and tildes, for example, $H_1$, $\tilde H_1$, $H_2$, etc. for integrals, and $h_1$, $\tilde h_1$, $h_2$, etc. for their Mellin images. It will be implied throughout that functions with the same index, $H_k$ and $h_k$, will be related by a transformation of the form \eqref{NfoldMBint}, or \eqref{2foldMBint} in the 2-fold case.

Now we can find the complete kernel $\bbW_\alpha\![\exp(-\tau F^\nu)/F^\mu | \sigma, m^2]$. To do this, it is enough to substitute the obtained function \eqref{AppEq2} into the integral transform \eqref{HybridizingTransform}:
\begin{align}
&\bbW_\alpha\!\Big[ \frac{e^{-\tau F^\nu}}{F^\mu} \Big| \sigma, m^2\Big] \nonumber \\
&\qquad= \frac{1}{\Gamma\left(\frac{\mu}{\nu}\right)} \int\limits_\tau^\infty ds\; (s-\tau)^{\frac{\mu}{\nu}-1} \bbW_\alpha\!\big[ e^{-\tau F^\nu} \big| \sigma, m^2\big] \nonumber  \\
&\qquad= m^{2\alpha}\int\limits_\tau^\infty ds\, \frac{(s-\tau)^{\frac{\mu}{\nu}-1}}{(4\pi)^{d/2}\Gamma\left(\frac{\mu}{\nu}\right)}  H_1\!\left(\frac{\sigma m^2}{2}, m^2 s^{1/\nu}\right) \nonumber  \\
&\qquad= \frac{m^{2\alpha} \tau^{\mu/\nu}}{(4\pi)^{d/2}}\; H_2\!\left(\frac{\sigma m^2}{2}, m^2 \tau^{1/\nu}\right),
\label{tildeKmunu}
\end{align}
where we have introduced a new function
\begin{equation} \label{H2Def}
H_2(z_1, z_2) = \int\limits_1^\infty dt\, \frac{(t-1)^{\frac{\mu}{\nu}-1}}{\Gamma\left(\frac{\mu}{\nu}\right)} \, H_1(z_1, z_2 t^{1/\nu}).
\end{equation}

Of course, we also want to represent the new function $H_2(z_1, z_2)$ as a 2-fold MB integral. To do this, it is sufficient to find the function $h_2(s_1, s_2)$, i.e. a double Mellin transform of~\eqref{H2Def} for which we have
\begin{align}
h_2(s_1, s_2)& = \int\limits_0^\infty dz_1\, z_1^{s_1-1} \int\limits_0^\infty dz_2\, z_2^{s_2-1} H_2(z_1, z_2) \nonumber \\
&= \frac{1}{\Gamma(\mu/\nu)}\,  h_1(s_1, s_2) \int\limits_1^\infty dt\, (t-1)^{\frac{\mu}{\nu}-1} t^{-\frac{s_2}{\nu}} \nonumber \\
&= \Gamma(s_1)\, \Gamma(s_1+s_2-\alpha)\, \frac{\Gamma\!\left(\frac{s_2-\mu}{\nu}\right)}{\nu\Gamma(s_2)}. \label{h2gamma}
\end{align}
Here we first substituted the formula \eqref{H2Def} into the transformation and changed the order of integration, then using a simple substitution we restored the image $h_1(s_1, s_2)$ and, finally, reduced the integral over $t$ to the Euler beta function and substituted the known expression for $h_1(s_1, s_2)$ \eqref{h1gamma}.

It is easy to see that introducing the factor $\hat F^{-\mu}$ into the operator function again, just as in the basis kernel case~\eqref{AppEq1}, leads to a shift of the argument of one of the gamma functions in \eqref{h2gamma} by $\mu$, and therefore, as it should be, limit $\mu=0$ restores the previously obtained result~\eqref{h1gamma}.

Here we first encounter another feature of our approach: the non-uniqueness of the multiple Mellin-Barnes representation for kernels. The arguments $z_1$, $z_2$ of the function $H_2(z_1,z_2)$~\eqref{H2Def} are dimensionless and are constructed from dimensional parameters of the complete kernel $\bbW_{\a}[F^{-\m}\exp(-\tau F^{\n})|\s,m^2]$, which are $\tau$, $m^2$, and $\s$. Obviously, the choice of dimensionless combinations is not unique, hence so is the multiple MB integral representation. The change of dimensionless arguments of $H_2$ leads to an alternative representation of~\eqref{tildeKmunu}, which, on the Mellin transform level corresponds to $\bbR$-linear transformations of the $\bm{s}$-space. Such alternative representations are useful since they allow to isolate a dimensional variable, such as $m^2$, which was previously present in both dimensionless arguments, inside a single dimensionless variable. This ``isolation'' is useful for studying various limits of expressions at hand. In this particular case substitution $s_1\mapsto s_1$, $s_2\mapsto s_2-s_1$ yields:
\begin{align}
&\bbW_\alpha\!\Big[ \frac{e^{-\tau F^\nu}}{F^\mu} \Big| \sigma, m^2\Big] = \frac{m^{2\alpha} \tau^{\mu/\nu}}{(4\pi)^{d/2}}\, \tilde H_2\!\Big(\frac{\sigma}{2\tau^{1/\nu}}, m^2\tau^{1/\nu}\Big), \label{tildeE1} \\
&\tilde h_2(s_1, s_2) = \Gamma(s_1) \Gamma(s_2-\alpha) \frac{\Gamma\left(\frac{s_2-s_1-\mu}{\nu}\right)}{\nu\Gamma(s_2-s_1)}. \label{tildeE3}
\end{align}
Notice, that, unlike in~\eqref{tildeKmunu}, $m^2$ is now present inside the second argument of the $\t H_2(z_1,z_2)$ only. It is in this form that we present these formulas in Appendix \eqref{E1App}-\eqref{E3App}.

To calculate various limits of \eqref{tildeE1} and similar functions of this type, which we will encounter later in this article, it is necessary to construct Horn series representations for a $N$-fold MB integral. Although this procedure is not particularly complicated, it requires an understanding of the general theory of $N$-fold MB integrals. Therefore, we will consider it separately in a upcoming accompanying paper \cite{BKW25b}. Here, for the reader's reference and convenience, we will limit ourselves to presenting all the limits without derivation in the Appendix \ref{SummaryAppendix}.

\subsection{Complete kernels for $1/(\hat F^\mu + \lambda)$}
\label{sec:hybrid_kernels}

We now consider the operator function $(\hat F^\mu + \lambda)^{-1}$. First of all, we need to obtain it from the previously found functions using some integral transform. This can be done in different ways---either using the integral transform \eqref{PowerToRes} or the Laplace transform \eqref{KernelToResolvent}:
\begin{align}
\frac{1}{\hat F^\mu + \lambda} &= \int\limits_C \frac{ds}{2\pi i}\, \Gamma(1-s)\Gamma(s) \lambda^{-s} \hat F^{\mu(s-1)} \label{IntResolvEq} \\
&= \int\limits_0^\infty dt\; e^{-t(\hat F^\mu + \lambda)}.
\end{align}
The first relation is already a Mellin--Barnes type integral, thus it is simpler, so we will use it. However, it can be directly verified that applying the second representation will produce the same results.

If one substitutes the basis $\bbB_\alpha\![F^{-\mu}| \sigma]$ \eqref{CompPowFunctions} or the complementary $\bbM_\alpha\![F^{-\mu}| m^2]$ \eqref{CompPowFunctions2} kernels into the \eqref{IntResolvEq} transform, one easily obtains the expressions given in Appendix~\ref{SummaryAppendix} for $\bbB_\alpha\![1/(F^\mu+\lambda)| \sigma]$ \eqref{calGEq1} and $\bbM_\alpha\![1/(F^\mu+\lambda)| m^2]$ \eqref{calGEq2}. Since these calculations do not contain anything fundamentally new, we will skip them to save space and immediately move on to considering the more complicated complete kernels $\bbW_\alpha\![1/(F^\mu+\lambda)| \sigma, m^2]$.

Substituting first the formula \eqref{GreenIRreg}, and then the representation for the Bessel--Clifford function \eqref{BCMeelinBarnes2}-\eqref{BCMeelinBarnes} into the integral transformation \eqref{IntResolvEq}, we obtain a representation of the desired complete kernel in the form of a 2-fold MB integral:
\begin{align}
&\bbW_\alpha\!\big[ (F^\mu + \lambda)^{-1} \big| \sigma, m^2\big] \nonumber \\
&= \int\limits_C \frac{ds}{2\pi i}\, \Gamma(1-s)\Gamma(s) \lambda^{-s} \bbW_\alpha\!\big[ F^{\mu(s-1)} \big| \sigma, m^2 \big] \nonumber \\
&= \frac{2m^{2(\alpha-\mu)}}{(4\pi)^{d/2}} \int\limits_C \frac{ds}{2\pi i} \frac{\Gamma(1-s)\Gamma(s)}{\Gamma(\mu(1-s))} \left(\frac{\lambda}{m^{2\mu}}\right)^{-s} \nonumber \\
&\,\times \calK_{\mu(s-1)+\alpha}\!\left(\frac{\sigma m^2}{2}\right) = \frac{m^{2(\alpha-\mu)}}{(4\pi)^{d/2}}\; H_3\!\left(\frac{\sigma m^2}{2}, \frac{\lambda}{m^{2\mu}}\right), \label{AppEq12}
\end{align}
where $H_3(z_1,z_2)$ is given by its Mellin image:
\begin{equation} \label{MlambdaEq2}
h_3(s_1, s_2) = \Gamma(s_1)\,\Gamma(s_2)\,\Gamma(1-s_2)\, \frac{\Gamma(s_1-\mu s_2+\mu-\alpha)}{\Gamma(\mu(1-s_2))}.
\end{equation}

Note that for $\mu=1$ the general relation \eqref{BHtransform} should lead to the following identity
\begin{equation}
\bbW_\alpha\big[(F+\lambda)^{-1}|\sigma, m^2\big] = \bbW_\alpha\big[F^{-1}|\sigma, m^2+\lambda\big].
\end{equation}
This identity is interesting because it relates the 2-fold MB integral to the 1-fold one
\begin{align}
H_3\!\Big(\frac{\sigma m^2}{2}, \frac{\lambda}{m^2}\Big)\Big|_{\mu=1} &= 2\Big(1+\frac{\lambda}{m^2}\Big)^{\alpha-1}\nonumber\\
&\times\calK_{\alpha-1}\!\left({\sigma(m^2+\lambda)}/{2}\right),
\end{align}
which is easy to check using the well-known relation
\begin{equation} \label{PowerOfSum}
(x+y)^{-\alpha} = \frac{x^{-\alpha}}{\Gamma(\alpha)} \int\limits_C \frac{ds}{2\pi i}\, \Gamma(-s)\Gamma(\alpha+s) \left(\frac{x}{y}\right)^{-s}.
\end{equation}

By changing variables $s_1\mapsto s_1+s_2$, $s_2\mapsto s_2/\mu$, we can rewrite the expressions \eqref{AppEq12}-\eqref{MlambdaEq2} in the following alternative form:
\begin{align}
&\bbW_\alpha\!\big[ (F^\mu + \lambda)^{-1} \big| \sigma, m^2\big] = \frac{m^{2(\alpha-\mu)}}{(4\pi)^{d/2}}\, \tilde H_3\!\Big(\frac{\sigma m^2}{2}, \frac{\sigma\lambda^{1/\mu}}{2}\Big), \label{MlambdaEq3} \\
&\tilde h_3(s_1, s_2) = \Gamma(s_1+\mu-\alpha)\,\Gamma(s_1+s_2)\, \frac{\Gamma(\tfrac{s_2}{\mu})\,\Gamma(1-\tfrac{s_2}{\mu})}{\mu\Gamma(\mu-s_2)}. \label{MlambdaEq4}
\end{align}
This is the form of notation given in Appendix \eqref{MlambdaEq3App}-\eqref{MlambdaEq4App}.

\subsection{The operator function $\exp(-\tau F^\nu)/(F^\mu + \lambda)$}

In this subsection we will consider the most complex function encountered in this article: $\exp(-\tau\hat F^\nu)/(\hat F^\mu + \lambda)$. In order to obtain the integral transform that generates it, it is sufficient to simply multiply the formula \eqref{IntResolvEq} by the exponential factor $e^{-\tau\hat F^\nu}$:
\begin{equation} \label{TheLastTransform}
\frac{e^{-\tau\hat F^\nu}}{\hat F^\mu + \lambda} = \int\limits_C \frac{ds}{2\pi i} \Gamma(1-s)\Gamma(s) \lambda^{-s} \frac{e^{-\tau\hat F^\nu}}{\hat F^{(1-s)\mu}},
\end{equation}
since the expressions for the hybrid function $\exp(-\tau\hat F^\nu)/\hat F^\mu$ are already known to us.

Accordingly, to calculate the basis kernel for this operator function, we must substitute into the transform \eqref{TheLastTransform} the expressions for the basis kernel $\bbB_\alpha\![e^{-\tau F^\nu}/F^\mu | \sigma]$ \eqref{bbK_mn_from_bbK_0n}, and then the representation for the function $\calE_{\nu,\alpha}^{(\mu)}(-z)$ as 1-fold MB integral \eqref{calKMellin}. As a result, we obtain
\begin{align}
&\bbB_\alpha\!\Big[ \frac{e^{-\tau F^\nu}}{F^\mu + \lambda} \Big| \sigma \Big] \nonumber \\
&\qquad= \int\limits_C \frac{ds}{2\pi i}\, \lambda^{-s} \Gamma(1-s)\Gamma(s)\, \bbB_\alpha\!\Big[ \frac{e^{-\tau F^\nu}}{F^{(1-s)\mu}} \Big| \sigma \Big] \nonumber \\
&\qquad= \frac{\tau^\frac{\mu-\alpha}{\nu}}{(4\pi)^{d/2}} \int\limits_C \frac{ds}{2\pi i}\, \Gamma(1-s)\Gamma(s) \nonumber \\
&\qquad\qquad\times \left(\lambda\tau^{\mu/\nu}\right)^{-s}\, \calE_{\nu,\alpha}^{((1-s_2)\mu)}\!\left(-\frac{\sigma}{2\tau^{1/\nu}}\right) \nonumber \\
&\qquad= \frac{\tau^\frac{\mu-\alpha}{\nu}}{(4\pi)^{d/2}}\; H_4\left(\frac{\sigma}{2\tau^{1/\nu}}, \lambda \tau^{\mu/\nu}\right), \label{HibridBasis1}
\end{align}
where
\begin{align}
&h_4(s_1, s_2) = \Gamma(s_2)\Gamma(1-s_2)\; \varepsilon_{\nu,\alpha}^{((1-s_2)\mu)}(s_1) \nonumber \\
&= \frac{\Gamma(s_1)\, \Gamma(s_2)\, \Gamma(1-s_2)}{\nu\Gamma(\alpha - s_1)} \Gamma\left(\frac{\mu s_2 - s_1 + \alpha - \mu}{\nu}\right). \label{HibridBasis3}
\end{align}
(Note, that the formulas \eqref{H4expressionApp}, \eqref{H4MBApp} were obtained via the change of variable $s_2\mapsto s_2/\mu$.)

First of all, we note that, as it should be, at $\mu=0$ the integrations become independent and the 2-fold integral splits into a product of two 1-fold ones: the integral over $s_1$ simply gives $(1+\lambda)^{-1}$, and the integral over $s_2$ again returns us to the generalized exponential function $\calE_{\nu,\alpha}(-z)$ \eqref{InvMellinCalE}-\eqref{MellinCalE}.

Again using linear transformations $s_1\mapsto s_1$, $s_2\mapsto s_2 + s_1/\mu$, the same result can be written in the equivalent form:
\begin{align}
&\bbB_\alpha\!\Big[ \frac{e^{-\tau F^\nu}}{F^\mu + \lambda} \Big| \sigma \Big] =  \frac{\tau^\frac{\mu-\alpha}{\nu}}{(4\pi)^{d/2}}\; \tilde H_4\left(\tfrac{1}{2}\sigma\lambda^{1/\mu}, \lambda \tau^{\mu/\nu} \right), \\
&\tilde h_4(s_1, s_2) = \frac{\Gamma(s_1)\, \Gamma(\tfrac{\mu s_2 + s_1}{\mu})\,\Gamma(1 - \tfrac{\mu s_2 + s_1}{\mu})}{\nu\Gamma(\alpha - s_1)}\nonumber\\
&\qquad\qquad\times\Gamma\left(\tfrac{\mu s_2 + \alpha - \mu}{\nu}\right).
\end{align}

Similarly, to obtain complete massive kernels, we must substitute the previously obtained complete kernels $\bbW_\alpha[\exp(-\tau F^\nu)/F^\mu | \sigma, m^2]$ \eqref{E1App} into the integral transform \eqref{TheLastTransform}. This leads us for the first time to an expression in which a 3-fold MB integral arises:
\begin{align}
&\bbW_\alpha\!\Big[ \frac{e^{-\tau F^\nu}}{F^\mu + \lambda} \Big| \sigma \Big] \nonumber \\
&= \int\limits_C \frac{ds}{2\pi i}\, \lambda^{-s} \Gamma(1-s)\Gamma(s)\, \bbW_\alpha\!\Big[ \frac{e^{-\tau F^\nu}}{F^{(1-s)\mu}} \Big| \sigma \Big] \nonumber \\
&= \frac{m^{2\a}\tau^{\mu/\nu}}{(4\pi)^{d/2}}\; H_5\left(\frac{\sigma}{2\tau^{1/\nu}}, m^2\tau^{1/\n}, \lambda^{1/\m} \tau^{1/\nu}\right), \label{HibridComplete1}
\end{align}
where
\begin{align}
h_5(s_1, s_2, s_3) &= \frac{\Gamma(s_1)\, \Gamma(s_2-\a)\, \,\Gamma(s_3/\m)\,\Gamma(1-s_3/\m)}{\mu\nu\,\Gamma(s_2 - s_1)}\nonumber\\
&\times\Gamma\left(\frac{s_2-s_1+s_3-\mu}{\nu}\right). \label{HibridComplete3}
\end{align}

\section{The resonant case} \label{ResonantCaseSec}

This final section is devoted to analysis of the consequences of treating the parameters $\alpha$, $\mu$, and $\nu$ as integers. Whereas before, having performed analytic continuation, we have assumed them to take some generic complex values, now it is time to substantiate our claims that Mellin--Barnes representations for basis kernels we have obtained above are meaningful for integer values of these parameters as well. Therefore, unless explicitly stated otherwise, in this section $\a$ is assumed to be integer, while $\m$ and $\n$ are assumed to take positive integer values.

In the Introduction we have stated a problem of infrared divergences of the proper-time integrals, which arise while performing the integral transform $\frakL_f$ used to obtain the basis kernel for $f(\h F)$. We have suggested two alternative ways of dealing with said divergences: regularization via analytic continuation and massive regularization, which amounts to calculating complete kernels $\bbW_{\a}\![f|\s,m^2]$ instead of $\bbB_{\a}\![f|\s]$. In what follows we consider the IR divergences in question which arise in basis kernels of the operator functions and show, that, in a certain sense, the two regularization procedures are equivalent: they produce identical finite and logarithmically divergent parts, whereas power divergences present in the massive regularization are invisible to the other. This is the well-known property of the dimensional regularization used in QFT \cite{Leibbrandt1975}.

In what follows, we will systematically encounter sets of the following form:
\begin{equation}
\bbZ_{\ge n} = \{k\in\bbZ \mid k\ge n\}, \quad \bbZ_{\le n} = \{k\in\bbZ \mid k\le n\},
\end{equation}
and use the corresponding notations. 

\subsection{Regularization via analytic continuation}

We start this section with considering the regularization via analytic continuation. As we have stated in preceding sections, the analytic continuation procedure does not fully eliminate IR divergences, but instead simply isolates true physical divergences as poles in the parameter $\alpha$ and allows one to work with them in a controlled manner. One can extract these poles explicitly by expanding the gamma function in the vicinity of the pole. This amounts to a shift $\alpha \mapsto \alpha + \epsilon$. Since $\alpha = d/2-k$, where $k\in\bbZ_{\ge0}$ and $d$ is the spacetime dimension, this shift has an obvious interpretation as the dimensional regularization.

\paragraph{Bare poles for $\hat F^{-\mu}$.}

In the simplest example of the basis kernel for complex power $\bbB_{\a}[F^{-\m}|\s]$ \eqref{CompPowFunctions}, the gamma function $\Gamma(\mu-\alpha)$ in the numerator has poles at $\mu - \alpha \in\bbZ_{\ge0}$. Therefore to obtain the regularized expression one uses the expansions for the complex power
\begin{equation} \label{PowerExpansion}
z^{-\epsilon} = e^{-\epsilon\ln z} = 1 - \epsilon\ln z + O(\epsilon^2),
\end{equation}
and the gamma function near the pole
\begin{equation} \label{GammaExpansion}
\Gamma(-n+\epsilon) = \frac{(-1)^n}{n!} \left( \frac{1}{\epsilon} + \psi(n+1)\right) + O(\epsilon),
\end{equation}
where $\psi(n+1) = H_n - \gamma$ is the digamma function, $H_n = \sum\nolimits_{k=1}^n k^{-1}$ are the harmonic numbers, and $\gamma$ is the Euler--Mascheroni constant. Substituting \eqref{PowerExpansion}-\eqref{GammaExpansion} into the formula \eqref{CompPowFunctions}, we get
\begin{align}
&\bbB_{\alpha+\epsilon}\!\lb F^{-\mu}|\sigma\rb = \frac{(-\s/2)^{\mu-\alpha}}{(4\pi)^{d/2}(\mu-1)!(\mu-\a)!} \nonumber \\
&\qquad\times\left( \frac{1}{\epsilon} - \ln\frac{\s}{2} + H_{\mu-\alpha} - \gamma \right) + O(\epsilon).
\label{gf_dim_reg}    
\end{align}

Note that, strictly speaking, akin to dimensional transmutation in dimensional regularization, we should introduce a dimensional parameter, which would restore the necessary dimensionality of the $\epsilon$-regularized basis kernel. This is precisely the parameter that would have made the argument of the logarithm in the second line of~\eqref{gf_dim_reg} dimensionless.

\paragraph{Pinches for $\exp(-\tau\hat F)/\hat F^\mu$.}

As we have seen in the previous paragraph, IR divergences are still present in the analytically continued basis kernels and they can be extracted via the dimensional regularization. So far, however, we have only looked at basis kernels which are known in terms of gamma functions, whereas the rest of basis kernels in this paper are derived in terms of Mellin--Barnes integrals. It turns out, that these IR divergences have an elegant interpretation in terms of MB integrals: they manifest themselves as contour pinches. Leaving a detailed explanation of this claim to the upcoming paper~\cite{BKW25b}, we only discuss this issue on the simplest of examples. 

In general, an integration contour of the $N$-fold MB integral should be chosen in such a way, that for each gamma functions in the numerator of the integrand, it leaves all its poles on one side. In the case $N=1$, there are only two possibilities: the poles of the gamma function can be either leftward-running (like in the case of $\Gamma(s)$), or rightward-running (as is the case for $\Gamma(\a-s)$). So, according to our prescription, the contour $C$ of the MB integral
\begin{equation}
    \int\limits_C\frac{ds}{2\pi i}z^{-s}\Gamma(s)\Gamma(\a-s)
\end{equation}
should pass in between the two rows of poles. For a generic values $\alpha\notin\bbZ_{\le0}$ this is always possible, but whenever $\alpha$ takes non-positive integer values, the contour $C$ becomes pinched between the poles of the two gamma functions.

As an example, consider the case of the basis kernel for the operator function $\exp(-\tau\hat F)/\hat F^\mu$ (i.e., $\nu=1$). According to the formula \eqref{bbBcalE}, we have
\begin{align}
&\bbB_{\alpha}\!\bigg[\frac{e^{-\tau F}}{F^\m}\bigg|\s\bigg] = \frac{\tau^{\m-\a}}{(4\pi)^{d/2}} \nonumber \\
&\qquad\times\int\limits_{C}\frac{ds}{2\pi i} \frac{\Gamma(s)\,\Gamma(\alpha-\mu-s)}{\Gamma(\alpha-s)} \lp\frac{\s}{2\tau}\rp^{-s}.
\end{align}
In the numerator of the integrand, the gamma function $\Gamma(s)$ generates a leftward-running row of poles $s=-k$, and the gamma function $\Gamma(\alpha-\mu-s)$ generates a rightward-running row of poles $s = \alpha-\mu+k$, where $k\in\bbZ_{\ge0}$. However, if $\mu\in\bbZ_{\ge1}$, almost all of these poles, except $\mu$, cancel out with the poles of the gamma function $\Gamma(\alpha-s)$ in the denominator according to the formula
\begin{equation}
\frac{\Gamma(\alpha-\mu-s)}{\Gamma(\alpha-s)} = \prod\limits_{k=1}^\mu (\alpha-k-s)^{-1}.
\end{equation}

Further, for $\alpha\in\bbZ_{\le\mu}$, some of these $\mu$ poles coincide with the poles of $\Gamma(s)$ at the points $\calP = \{\alpha-\mu, \ldots, n_\alpha\}$, where $n_\alpha = \min\{\alpha-1, 0\}$. So the integration contour $C$ is pinched between poles in $\calP$. There are $\mu-\alpha+1$ pinches for $1\le\alpha\le\mu$ and exactly $\mu$ pinches for $\alpha\in\bbZ_{\le1}$. Regularization of the corresponding expression using the above-described shift $\alpha \mapsto \alpha + \epsilon$ leads to the following series
\begin{equation} \label{hybrid_m1_reg}
\bbB_{\a+\epsilon}\!\bigg[\frac{e^{-\tau F}}{F^\m}\bigg|\s\bigg] = \frac{\tau^{\m-\a}}{(4\pi)^{d/2}}\sum\limits_{n=0}^{\infty}\frac{c_n(\epsilon)}{n!}\lp-\frac{\s}{2\tau}\rp^n,
\end{equation}
where
\begin{equation}
c_n(\epsilon) = \tau^{-\epsilon} \prod\limits_{k=1}^\mu (n+\alpha+\epsilon-k)^{-1}.
\end{equation}
Obviously, poles $-n\notin\calP$ where there are no pinches give finite contributions to this sum, $c_n(\epsilon) \xrightarrow[\epsilon\to0]{} c_n(0)$. Meanwhile, contributions from poles $-n\in\calP$ where pinches occur contain divergences. Using the expansion \eqref{PowerExpansion}, we obtain:
\begin{equation} \label{CnEpsilon}
c_n(\epsilon) = \left(\frac{1}{\epsilon} - \ln\tau\right) \prod\limits_{\tln{k=1}{k\ne n+\alpha}}^{\m} (n+\alpha-k)^{-1} + O(\epsilon).
\end{equation}

In order to further compare these results with the formulas obtained by regularization via the mass parameter, we present separately the expressions for the simplest case $\mu=1$. Note that then the answer can be rewritten as
\begin{equation} \label{Bmu1Eq}
\bbB_\alpha\!\bigg[ \frac{e^{-\tau F}}{F} \bigg| \sigma \bigg] = \frac{(\sigma/2)^{1-\alpha}}{(4\pi)^{d/2}}\; \gamma\!\left(\alpha-1, \frac{\sigma}{2\tau}\right),
\end{equation}
were
\begin{align}
\gamma(\alpha, z) &= \int\limits_0^z t^{\alpha-1}dt\, e^{-t} = z^\alpha\sum\limits_{k=0}^\infty \frac{(-z)^k}{(\alpha+k) k!} \nonumber \\
&= z^\alpha \int\limits_C \frac{ds}{2\pi i}\frac{\Gamma(s)}{\alpha-s}\, z^{-s} \label{IncompleteGammaExpansion}
\end{align}
is the lower incomplete gamma function. The integral in the first line converges for $\Re\alpha>0$, while the representation as the MB integral in the second line, where the integration contour $C$ passes between the pole $s=\alpha$ and the poles of the function $\Gamma(s)$, provides an analytic continuation of this function into the parameter domain $\Re\alpha\le0$, with the exception of the points $\alpha\in\bbZ_{\le0}$, at which pinches occur and the function $\gamma(\alpha, z)$ has singularities. In the resonant case $\alpha\in\bbZ_{\le1}$, we obtain the following dimensionally regularized expression:
\begin{align}
&(4\pi)^{d/2}\, \bbB_{\a+\epsilon}\!\bigg[\frac{e^{-\tau F}}{F}\bigg|\s\bigg] = \frac{(-\s/2)^{1-\a}}{(1-\alpha)!} \left(\frac{1}{\epsilon}-\ln\tau\right) \nonumber \\
&\qquad+ \tau^{1-\a}\!\! \sum\limits_{\tln{n=0}{n\neq1-\a}}^{\infty} \frac{(-\sigma/2\tau)^n}{(n+\a-1) \,n!}+O(\epsilon). \label{massless_hybrid_1_1_limit_1}
\end{align}

Finally, we note that in the case $\mu=0$ and $\nu\in\bbZ_{\ge1}$ in formula \eqref{AppEq1}, all the right-running poles $s = \alpha+k\nu$ of the gamma function $\Gamma\big((\alpha-s)/\nu\big)$ in the numerator cancel out without exception with the poles $s=\alpha+k$ of the gamma function $\Gamma(\alpha-s)$ in the denominator. Since there are no rightward-running poles at all in this case, pinches and associated divergences cannot arise. This means that the basis kernels for the operator exponentials $\exp(-\tau\hat F^\nu)$ for $\nu\in\bbZ_{\ge1}$ are well-defined functions for any values of the parameter $\alpha$, as we have already indicated earlier in \cite{Wach2}.

\subsection{Regularization via mass introduction}

We now turn to the consideration of the complete kernels $\bbW_\alpha[f | \sigma, m^2]$ \eqref{BHtransform}, which arise when the mass term $m^2$ in the operator is taken into account ``nonperturbatively.'' As we discussed in the Introduction, in this case $m^2$ serves as an IR regularizer, resulting in all integrals converging well, and there are no any divergences at all. However, they can obviously reappear in the massless limit $m\to0$. The question arises whether this limit reproduces the results obtained by the alternative method of analytic continuation, and if so, in what sense.

\paragraph{Complete kernels for $\hat F^{-\mu}$.}

Let us start with the simplest case of a complete massive kernel for the operator complex power $\hat F^{-\mu}$, $\mu\in\bbZ_{\ge1}$. The corresponding expression in terms of the Bessel--Clifford function is given by the formula \eqref{GreenIRreg}. We must substitute into this expression the asymptotics of $\calK_N(z)$ \eqref{BC_resonant_pos} for the resonant case $N = \alpha-\mu$, $\alpha\in\bbZ$. Since we are interested in the massless limit $m^2\to0$, we will discard terms of positive power in $m^2$ which vanish in it.

Thus, in the relevant region $\alpha\in\bbZ_{>\mu}$, the answer will be given by the single term of the lowest power $z^{-N}$ in the formula \eqref{BC_resonant_pos}
\begin{equation} \label{massive_power_1_1_limit_3}
\bbW_\alpha\!\lb F^{-\mu}|\s,m^2\rb = \frac{(\alpha-\mu-1)!}{(4\pi)^{d/2} (\mu-1)!} \left(\frac{\sigma}{2}\right)^{\mu-\alpha} + O(m^2).
\end{equation}
It does not contain any divergences and exactly coincides with the answer \eqref{CompPowFunctions} given by analytic continuation. Of course, this is as it should be, since the relevant region is characterized by the fact that no IR divergences arose in it.

The situation in the irrelevant region $\alpha\in\bbZ_{\le\mu}$ will be less trivial. In this case, we must transform the asymptotics of \eqref{BC_resonant_pos} using the relation \eqref{MinusAlpha}. Substituting it into the formula \eqref{GreenIRreg}, we get
\begin{align}
&\bbW_\alpha\!\lb F^{-\mu}|\s,m^2\rb = \frac{(-\sigma/2)^{\mu-\alpha}}{(4\pi)^{d/2} (\mu-1)!} \nonumber \\
&\qquad\times\bigg( \sum\limits_{n=1}^{\mu-\alpha} \frac{(n-1)!}{(\mu-\alpha-n)!} \lp-\frac{\sigma m^2}{2}\rp^{-n} \nonumber \\
&\qquad-\frac{\ln\frac{\s m^2}{2} + 2\gamma - H_{\mu-\alpha}}{(\mu-\alpha)!} \bigg) + O(m^2). \label{massive_power_1_1_limit_1}
\end{align}
This result differs from the formula \eqref{gf_dim_reg} by the appearance of $(\mu-\alpha)$ additional terms with negative powers of $m^2$ (they are absent for the marginal case $\alpha=\mu$). However, the contributions which logarithmic and constant in $\sigma$ exactly coincide if we set
\begin{equation} \label{EpsilonLogGamma}
 -\frac{1}{\epsilon} = \ln m^2 + \gamma.
\end{equation}
This situation is well known in QFT: the dimensional regularization only captures the logarithmically divergent parts of the integrals, while all power divergences vanish identically.

Note that the cases considered are also resonant, although pinches and the divergences they generate do not arise in them. This is because, while in the case of pinches, the coalescing simple poles are on opposite sides of the integration contour, in this case, simple poles lying on the same side of the integration contour coalesce, leading to the emergence of higher-order poles and associated logarithmic terms. Therefore, the corresponding case is usually called \emph{logarithmic} in the literature.

\paragraph{Complete kernels for $\exp(-\tau\hat F)/F$.}

A similar situation can be observed for an object, whose massless counterpart experienced an MB contour pinch---the $\m=\n=1$ complete hybrid basis kernel $\bbW_\alpha[ \exp(-\tau F)/F | \s,m^2 ]$. It is given by a 2-fold Mellin-Barnes integral~(\ref{tildeE3}), so, in some sense, the introduction of the mass parameter is also a pinch regularization procedure as it allows one to pass the integration contour in a way that avoids pinch by making the complex integration domain 2-dimensional.

The derivations of resonant series representations are quite lengthy, require the use of a special technique, and are thus relegated to the accompanying paper~\cite{BKW25b}. Here, we present only the final answers without proof, and, as above, we discard terms with positive powers of $m^2$ (since we are interested in the massless limit of $m^2\to0$).

For the relevant region $\alpha\in\bbZ_{>1}$, we obtain that, up to higher-order corrections in $m^2$, the complete kernel coincides with the previously obtained basis kernel \eqref{Bmu1Eq}
\begin{equation} \label{massive_hybrid_1_1_limit_3}
\bbW_\alpha\!\bigg[\frac{e^{-\tau F}}{F} \bigg | \sigma, m^2\bigg] = \bbB_\alpha\!\bigg[\frac{e^{-\tau F}}{F} \bigg | \sigma \bigg] +O(m^2).
\end{equation}
Again, this is exactly what one would expect, since there were no initially IR divergences in the relevant region.

In the nontrivial irrelevant region $\alpha\in\bbZ_{\le1}$, where the massless limit contains IR divergences, the answer is more complicated:
\begin{align}
&(4\pi)^{d/2}\bbW_\alpha\!\bigg[ \frac{e^{-\tau F}}{F} \bigg|\sigma, m^2\bigg] \nonumber \\
    &\qquad= m^{2(\alpha-1)} \sum\limits_{n=0}^{-\a}\frac{(-\a-n)!}{n!} \lp-\frac{\sigma m^2}{2}\rp^n \nonumber \\
    &\qquad+ \tau^{1-\a}\!\! \sum\limits_{\tln{n=0}{n\neq1-\a}}^{\infty}\frac{(\sigma/2\tau)^n}{(n+\a-1)n!} \nonumber \\
    &\qquad-\frac{(-\s/2)^{1-\a}}{(1-\a)!} \big( \ln(\tau m^2) + \gamma \big) + O(m^2), \label{massive_hybrid_1_1_limit_1}
\end{align}
Comparing this result with the previously obtained expression for the basis kernel \eqref{massless_hybrid_1_1_limit_1}, we again find that we additionally have $(1-\alpha)$ terms with negative powers of $m^2$, diverging in the massless limit (the second line, they vanish in the marginal case $\alpha=1$). If we consider the remaining part, logarithmic or independent of $m^2$ (the third and fourth lines), it exactly coincides with \eqref{massless_hybrid_1_1_limit_1}, assuming that the relation \eqref{EpsilonLogGamma} holds.

The appearance of additional terms of negative power in $m^2$ can be easily explained using the ideas we developed in \cite{BKWLetter}. As we explained there, the expansions of integral kernels of operator functions include terms coming from either the UV or the IR domain. Term-by-term integration of the DeWitt series followed by analytic continuation captures exactly all UV terms and knows nothing about the IR terms. However, when we ``nonperturbatively'' take into account the mass term $m^2$ by going to complete massive kernels $\bbW_\alpha$, the answer also includes some IR terms accounting for the exponentially decaying factor $\exp(-\tau m^2)$. Moreover, in the irrelevant region, these terms will include those that diverge in the massless limit $m^2\to0$. These are precisely the terms we see in the formula \eqref{massive_power_1_1_limit_1} compared to \eqref{gf_dim_reg}, and in the formula \eqref{massive_hybrid_1_1_limit_1} compared to \eqref{massless_hybrid_1_1_limit_1}.

The following two ideas are key for us: the first is the above-described notion of UV and IR contributions, which are completely independent in the regularized non-resonant case. The second is that the answers for the resonant case can be obtained by the limit from the non-resonant case. Although this causes the UV and IR contributions to interact non-trivially, the checks performed in this section confirm that these two ideas are compatible.

\subsection{Cancellation of IR divergences}

There is one more notable thing regarding the hybrid kernels---the apparent finiteness of certain linear combinations thereof with the kernels of complex powers. This stems from the fact that these combinations can be obtained using proper time integrals with finite limits~\cite{Barvinsky25}. In this subsection we show that these linear combinations are finite in our approach, both within the mass- and massless-pinch regularizations, which further solidifies the trust in the consistency of our two approaches.

The simplest example of the finite linear combination described above arises, for example, both in the Proca heat kernel~\cite{BarvinskyKalugin2024} and in the heat kernel of the nondegenerate nonminimal vector operator \cite{Barvinsky25}, which both contain derivatives of structures of the type $\big(e^{-\tau \h F}-1\big)/\hat F$. These structures can be obtained by means of proper time integral with finite limits:
\begin{equation} \label{dif_1_1}
    \frac{e^{-\tau \h F}-1}{\h F}=-\int\limits_0^{\tau}dt\,e^{-t \h F}
\end{equation}
and therefore are expected to be well-behaved in the IR. 

\paragraph{Analytic regularization.}

Note that by combining the formulas \eqref{CompPowFunctions} and \eqref{Bmu1Eq}, we can write
\begin{equation} \label{incompleteGdifference}
\bbB_{\a}\!\bigg[\frac{e^{-\tau F}-1}{F}\bigg|\s\bigg] = -\frac{(\s/2)^{1-\a}}{(4\pi)^{d/2}}\; \Gamma\! \lp \a-1,\frac{\s}{2\tau} \rp,
\end{equation}
where
\begin{equation} \label{UpperGamma}
\Gamma(\alpha, z) = \int\limits_z^\infty t^{\alpha-1} e^{-t} dt = \Gamma(\alpha) - \gamma(\alpha, z)
\end{equation}
is the upper incomplete gamma function. Of course, \eqref{incompleteGdifference} can also be obtained by termwise integrating the DeWitt expansion using \eqref{dif_1_1}, as was done in \cite{Barvinsky25}.

Although in \eqref{UpperGamma} $\Gamma(\alpha)$ and $\gamma(\alpha, z)$ have singularities at $\alpha\in\bbZ_{\le0}$, the integral converges absolutely for any $\alpha$ and $z>0$. Therefore, $\Gamma(\alpha, z)$ is an entire function of the parameter $\alpha$, and the singularities of $\Gamma(\alpha)$ and $\gamma(\alpha, z)$ cancel each other out.

Exactly the same cancellation of IR divergences occurs in the formula \eqref{incompleteGdifference}. This can be verified directly for $\alpha\in\bbZ_{\le1}$ by combining the formulas \eqref{gf_dim_reg} and \eqref{massless_hybrid_1_1_limit_1}. The divergent terms $\sim1/\epsilon$ in them are exactly the same, and after their cancellation we obtain:
\begin{align}
    &(4\pi)^{d/2}\, \bbB_{\a+\epsilon}\!\bigg[\frac{e^{-\tau F} - 1}{F}\bigg|\s\bigg] \nonumber \\
&\qquad= \frac{(-\s/2)^{1-\a}}{(1-\a)!} \Big( \ln\frac{\s}{2\tau} + \gamma - H_{1-\alpha} \Big) \nonumber \\
&\qquad+ \tau^{1-\a}\!\! \sum\limits_{\tln{n=0}{n\neq1-\a}}^{\infty}\frac{(-\sigma/2\tau)^n}{(n+\a-1) \,n!}+O(\epsilon), \label{massless_subtraction_1_1_answer}
\end{align}

There is an analog of the identity~\eqref{dif_1_1} for $\m\in\bbZ_{>1}$. The corresponding finite linear combination appears in nonminimal heat kernels is given by $\m$-fold analogue of the proper time integral \eqref{dif_1_1} which reads~\cite{Barvinsky25}
\begin{equation}
\begin{split}
&(-1)^{\m}\int\limits_0^{\tau}dt_1\int\limits_0^{t_1}dt_2\cdots\int\limits_0^{t_{\m-1}}dt_\m\,e^{-t_{\m}\h F}\\
&\qquad=\frac{1}{\h F^{\m}}\lb e^{-\tau \h F}-\sum\limits_{n=0}^{\m-1} \frac{(-\tau\h F)^{n}}{n!} \rb.
\end{split}
\label{m1_ir_fin_difference}
\end{equation}

Using \eqref{gf_dim_reg} and \eqref{hybrid_m1_reg}-\eqref{CnEpsilon} one sees that in the combination
\begin{equation}
    \bbB_{\a+\epsilon}\!\bigg[\frac{e^{-\tau F}}{F^{\m}}\bigg|\s\bigg]-\sum_{n=0}^{\m-1}\frac{(-\tau)^n}{n!}\bbB_{\a+\epsilon}\!\lb F^{\m-n}|\s\rb = O(\epsilon^0)
\end{equation}
each of the $\mu$ terms cancels exactly one divergence associated with the $\mu$ pinches in $\bbB_{\alpha+\epsilon}[\exp(-\tau F)/F^\mu | \sigma]$ \eqref{hybrid_m1_reg}. Therefore, the entire expression turns out to be finite at $\epsilon\to0$ as expected.

\paragraph{Massive regularization.}

Since we have already established that the two regularizations coincide in the finite and logarithmically divergent parts, it is no surprise that the massive regularization results in the cancellation of logarithmically divergent parts in the combinations like~\eqref{dif_1_1}. But so far nothing could be said about the power-divergent parts of the corresponding complete kernels. A straightforward calculation show that the power divergences in \eqref{massive_hybrid_1_1_limit_1}, just as the logarithmic ones, are subtracted by the corresponding terms in the massive Green function's basis kernels, so the combined object~(\ref{dif_1_1}) is devoid of any singularities at $m^2\to 0$ and coincides with the one obtained in the massless case. 

So, for example, in the case $\mu=1$, $\alpha\in\bbZ_{\le1}$, subtracting \eqref{massive_power_1_1_limit_1} from \eqref{massive_hybrid_1_1_limit_1}, we obtain that the complete massive kernel, up to terms of positive degree in $m^2$, coincides with the basis kernel
\begin{equation} \label{incompleteGdifference_massive}
\bbW_\alpha\!\bigg[\frac{e^{-\tau F}-1}{F}\bigg|\sigma, m^2\bigg] = \bbB_\alpha\!\bigg[\frac{e^{-\tau F}-1}{F}\bigg|\sigma \bigg] + O(m^2),
\end{equation}
which once again illustrates the legitimacy of the pinch $\epsilon$-regularization procedure and our technique of analytic continuation in the parameter $\alpha$.

\section{Conclusion} \label{sec:conclusion}

In this paper, we continue our study of integral kernels $f(\hat F)\delta(x,x')$ \eqref{pdo_kernel} for minimal second-order differential operators $\hat F(\nabla)$ on a curved background, which are important for applications. As we explained in \cite{BKWLetter}, these objects can be expanded in the form of a functional series containing terms coming from the UV and IR regions. Moreover, the UV terms can be obtained by term-by-term integration of the DeWitt series \eqref{HeatKernelExpansion}. In this case, IR divergences arise in the integrals, which must be regularized in some way. We consider two types of regularization: the regularization via analytic continuation allows us to obtain exactly all UV terms and leads to objects that we call basis kernels $\bbB_\alpha[f | \sigma]$ \eqref{frakLtransformEq2}. Regularization using the nonperturbative account of the mass term $m^2$ leads to slightly more complex complete massive kernels $\bbW_\alpha[f | \sigma, m^2]$ \eqref{BHtransform}, which include not only UV terms but also some IR terms.

Remarkably, this procedure allows us to separate two types of data: in the \eqref{general_kernel_series_rep} or \eqref{ModifiedExpansion} expansions, all information about the bundle geometry and the specific form of the operator $\hat F(\nabla)$ is encoded in the known HaMiDeW coefficients $\hat a_n[F | x,x']$, while the form of the basis $\bbB_\alpha[f | \sigma]$ and complete $\bbW_\alpha[f | \sigma, m^2]$ kernels---analytic functions of the parameters $\alpha$, $\sigma$, and $m^2$---are determined solely by the functional dependence $f$. This phenomenon is an off-diagonal generalization of the property called ``funtoriality'' in the literature; therefore, we call it ``off-diagonal functoriality.''

In  sections \ref{MainSection} and \ref{ResultsSection}, we presented several concrete examples of this approach, calculating the basis $\bbB_\alpha$ and complete $\bbW_\alpha$ kernels for several practically important functions $f$. It turns out that all these objects can be represented in a universal form using $N$-fold MB integrals. This fact, together with the integral transform technique we employ, makes the approach we develop particularly convenient, flexible, and powerful. Some further capabilities of our method will be demonstrated with specific examples in an upcoming article \cite{BKW25b}.

We use dimensional regularization throughout. For almost all values of the parameter $\alpha = d/2 - k$, the results obtained pertain to the so-called non-resonant case, where the UV and IR contributions do not interact with each other, and all functions have representations as Horn power series. However, for certain distinguished values of $\alpha$---and often when dimensional regularization is removed---the so-called resonant case arises, where the UV and IR contributions begin to interact, and the corresponding poles coalesce. This leads either to the appearance of pinches of integration contours, which we interpret as true physical IR divergences, or to the emergence of higher-order poles and associated logarithmic terms in the expansion. Therefore, in section \ref{ResonantCaseSec}, we consider in detail a number of examples related to the resonant case and show that our method works well and yields consistent results in this important case as well.

Of the possible directions in which our approach can be further developed, we point out two that we believe are the most interesting. First, our assertion about UV and IR contributions in the expansions of integral kernels requires additional verification from the infrared side. This requires studying the infrared asymptotics of the heat kernel $\hat K_F(\tau | x,x')$ as $\tau\to\infty$, which is a much more subtle problem. Therefore, it is natural to begin such a verification with the case of relatively simple manifolds: hyperbolic spaces and hyperspheres (although in the latter case, the exact answer is not expressed in terms of MB integrals, requiring more complex functions of elliptic type), as well as asymptotically flat spaces. The second practically important direction is the application of our off-diagonal method to the calculation of the Wodzicki residue and to the study of anomalies in the functional determinants.

\section*{Acknowledgments}
 The work of AOB and AEK was supported by the grant from the ``BASIS'' Foundation for the Advancement of Theoretical Physics and Mathematics.


\appendix
\section{The Bessel--Clifford function $\calK_\alpha(z)$} \label{BesselCliffordAppendix}

In this appendix we consider an important example of the Bessel--Clifford functions of the second kind in detail. Since these functions play a crucial role in our entire study, we provide the reader with the series representations of these functions as well as with an easy way to derive them with the help of the Mellin-Barnes integral, which is a central object of our paper.

The Bessel--Clifford functions of the second kind can be defined as
\begin{equation} \label{BC2Def}
\calK_\alpha(z) = \frac{1}{2} \int\limits_0^\infty dt\; t^{-\alpha-1} \exp\left(-t - \frac{z}{t}\right).
\end{equation}
The substitution $t\mapsto z/t$ leads to the relation
\begin{equation} \label{MinusAlpha}
\calK_{-\alpha}(z) = z^\alpha \calK_\alpha(z).
\end{equation}
Note that the well-known MacDonald functions $K_\alpha(z)$ are expressed in terms of $\calK_\alpha(z)$ as follows
\begin{equation} \label{McDonaldDef}
K_\alpha(z) = \Big(\frac{z}{2}\Big)^\alpha\; \calK_\alpha\!\Big(\frac{z^2}{4}\Big).
\end{equation}

As we explained in \cite{BKWLetter}, the representation \eqref{BC2Def} can serve as a toy model of our general idea of term-by-term integration of the DeWitt series. In this analogy, variable $t$ corresponds to the proper time $\tau$, the integrand $\exp(-t-z/t)$ is the heat kernel $\hat K_F(\tau | x, x')$, the integral over $t$ will be the integral transform $\frakL_f$ \eqref{frakLtransformEq}, and the function $\calK_\alpha(z)$ itself will serve as the kernel $\hat K[f(F) | x, x']$.

We want to study the behavior of the function $\calK_\alpha(z)$ \eqref{BC2Def} in the ``coincidence limit'' $z\to0$. Our term-by-term integration logic amounts to expanding the factor $e^{-t}$ in the integrand into a power series in $t$, then swapping summation and integration, and calculating the integrals with help of the Euler integral $\Gamma(z) = \int\nolimits_0^\infty t^{z-1} e^{-t} dt$. This yields the following ``UV'' expansion:
\begin{align}
\calK_\alpha^\mathrm{UV}(z) &= \frac{1}{2} \sum\limits_{k=0}^\infty \frac{(-1)^k}{k!} \int\limits_0^\infty dt\; t^{k-\alpha-1} e^{-z/t} \nonumber \\
&= \frac{z^{-\alpha}}{2} \sum\limits_{k=0}^\infty \Gamma(\alpha-k) \frac{(-z)^k}{k!}.  \label{BCseries2}
\end{align}

On the other hand, we can perform a similar procedure in the opposite, ``IR'' region. Expansion of the factor $e^{-z/t}$ into a series and swapping summation and integration yields a completely different expansion:
\begin{align}
\calK_\alpha^\mathrm{IR}(z) &= \frac{1}{2} \sum\limits_{k=0}^\infty \frac{(-z)^k}{k!} \int\limits_0^\infty dt\; t^{-\alpha-k-1} e^{-t} \nonumber \\
&= \frac{1}{2} \sum\limits_{k=0}^\infty \Gamma(-\alpha-k) \frac{(-z)^k}{k!}. \label{BCseries1}
\end{align}

The fact that the ``UV'' \eqref{BCseries2} and ``IR'' \eqref{BCseries1} expansions do not coincide with each other $\calK_\alpha^\mathrm{UV}(z) \ne \calK_\alpha^\mathrm{IR}(z)$ is not surprising: the point is that at least one of the two integrals always diverges. The integral in \eqref{BCseries1} diverges at the lower limit $t=0$ for $\Re\alpha>-k$, and the integral in \eqref{BCseries2} diverges at the upper limit $t=\infty$ for $\Re\alpha<k$. Therefore, the condition of the Fubini--Tonelli theorem is not satisfied, and the trick with swapping summation and integration, strictly speaking, does not work.

However, what is truly remarkable and worthy of attention here is something entirely different: in fact, the correct asymptotics of the Bessel--Clifford function $\calK_\alpha(z)$ \eqref{BC2Def} for the ``non-resonant'' case $\alpha\notin\bbZ$ is given by the sum of these ``UV'' and ``IR'' contributions (for the ``resonant'' case $\alpha\in\bbZ$ this is true in some limit sense, as will be explained below):
\begin{equation} \label{BCasymptotic}
\calK_\alpha(z) = \calK_\alpha^\mathrm{IR}(z) + \calK_\alpha^\mathrm{UV}(z).
\end{equation}
To see this, one finds the Mellin transform of the Bessel--Clifford function:
\begin{align}
\kappa_\alpha(s) &= \int\limits_0^\infty \calK_\alpha(z) z^{s-1}dz \nonumber \\
&= \frac{1}{2} \int\limits_0^\infty dt\; t^{-\alpha-1} e^{-t} \int\limits_0^\infty dz\; z^{s-1} e^{-z/t} \nonumber \\
&= \frac{\Gamma(s)}{2} \int\limits_0^\infty dt\; t^{s-\alpha-1} e^{-t} = \frac{1}{2}\Gamma(s)\Gamma(s-\alpha). \label{BCMeelinBarnes2}
\end{align}
Then the inverse Mellin transform gives the representation of the function $\calK_\a(z)$ as the Mellin--Barnes integral
\begin{equation} \label{BCMeelinBarnes}
\calK_\alpha(z) = \int\limits_C \frac{ds}{2\pi i}\, z^{-s} \kappa_\alpha(s).
\end{equation}
The substitution $s\mapsto s-\alpha$ again leads to the relation \eqref{MinusAlpha}. Closure of the integration contour on the right reduces \eqref{BCMeelinBarnes} to the sum of the residues at the poles of $\kappa_\alpha(s)$.

In the non-resonant case $\alpha\notin\bbZ$, the poles of the two gamma functions in \eqref{BCMeelinBarnes2} do not coincide with each other, so the residues in them can be taken independently. In this case, the sum of the residues at the poles of $\Gamma(s-\alpha)$ is exactly equal to the contribution $\calK_\alpha^\mathrm{UV}(z)$ \eqref{BCseries2}, while the sum of the residues at the poles of $\Gamma(s)$ is equal to the contribution $\calK_\alpha^\mathrm{IR}(z)$ \eqref{BCseries1}. We can say that the ``UV'' and ``IR'' contributions do not interact with each other in this case. If we are interested only in the leading term of the asymptotic expansion of $\calK_\alpha(z)$ as $z\to0$, then depending on the sign of $\Re\alpha$ it will come from either the ``UV'' or the ``IR'' part of the expansion \eqref{BCasymptotic}. Then we obtain the following analogue of the Stokes phenomenon:
\begin{equation}
\calK_\alpha(z) \approx \begin{cases}
z^{-\alpha}\, \Gamma(\alpha)/2, & \quad\text{for}\quad \Re\alpha>0, \\
\Gamma(-\alpha)/2, & \quad\text{for}\quad \Re\alpha<0.
\end{cases} \label{BCasymptotic}
\end{equation}

In the resonant case, the situation is only slightly more complicated: the simple poles of the gamma functions $\Gamma(s)$ and $\Gamma(s-\alpha)$ coalesce, i.e., the ``UV'' and ``IR'' contributions begin to interact with each other. Therefore, to obtain the expansion, we only need to carefully calculate the residues at the resulting double poles, which will lead to the appearance of the logarithms $\ln z$ and the Euler--Mascheroni constant $\gamma$ and the harmonic numbers $H_n$ in coefficients. More specifically, for $\alpha = N = 0, 1, 2, \ldots$ we obtain the following expansion:
\begin{align}
    &\calK_N(z) = \frac{(-1)^N}{2} \bigg( \sum\limits_{n=1}^N \frac{(n-1)!}{(N-n)!}\; (-z)^{-n} \nonumber \\
    &\quad-\sum\limits_{n=0}^{\infty}\frac{\ln z + 2\gamma - H_n - H_{N+n}}{n!(n+N)!}\; z^n \bigg), \label{BC_resonant_pos}
\end{align}
where the first sum consists of $N$ residues at the simple poles (and vanishes at $N=0$), and the second sum consists of an infinite number of residues at the double poles. For $\alpha = -N$, the corresponding answer is given by applying the relation \eqref{MinusAlpha}. Obviously, the expansion \eqref{BC_resonant_pos} for the resonant case can be obtained from the expansion \eqref{BCasymptotic} for the non-resonant case by taking the limit $\alpha\to N$.

It is interesting to note the following: let us try to correct the mistake in our initial naive reasoning by breaking the original integral \eqref{BC2Def} into two---from 0 to some positive constant $a$ and from $a$ to $\infty$. We expand the factor $e^{-t}$ in the first integral, and the factor $e^{-z/t}$ in the second integral. In this case, all integrals are well defined, the conditions of the Fubini--Tonelli theorem are satisfied, and we arrive at the following correct expansion
\begin{equation} \label{IncompleteExpansion}
\calK_\alpha(z) = \tilde\calK_\alpha^\mathrm{IR}(z, a) + \tilde\calK_\alpha^\mathrm{UV}(z, a),
\end{equation}
where 
\begin{align}
\tilde\calK_\alpha^\mathrm{IR}(z, a) &= \frac{1}{2} \sum\limits_{k=0}^\infty \Gamma(-\alpha-k, a) \frac{(-z)^k}{k!}, \\
\tilde\calK_\alpha^\mathrm{UV}(z, a) &= \frac{z^{-\alpha}}{2} \sum\limits_{k=0}^\infty \Gamma(\alpha-k, \tfrac{z}{a}) \frac{(-z)^k}{k!},
\end{align}
and $\Gamma(s,x) = \Gamma(s) - \gamma(s, x)$ is the upper incomplete gamma function \eqref{UpperGamma}. If we substitute into the formula \eqref{IncompleteExpansion} the known series expansion for $\gamma(s, x)$ \eqref{IncompleteGammaExpansion}, then it is easy to see that all its terms cancel out, and we will again return to \eqref{BCasymptotic}. Thus, the decomposition \eqref{IncompleteExpansion} corresponds to some mixing of the ``UV'' and ``IR'' contributions.

However, we cannot pass to the limits $a\to0$ or $a\to\infty$ in \eqref{IncompleteExpansion}. Indeed, in the limit $a\to0$ the term $\tilde\calK_\alpha^\mathrm{UV}(z, a)$ vanishes, and \eqref{IncompleteExpansion} reduces to $\calK_\alpha^\mathrm{IR}(z)$ \eqref{BCseries1}. Conversely, in the opposite limit $a\to\infty$ the term $\tilde\calK_\alpha^\mathrm{IR}(z, a)$ vanishes, and \eqref{IncompleteExpansion} reduces to $\calK_\alpha^\mathrm{UV}(z)$ \eqref{BCseries2}. Thus, \eqref{IncompleteExpansion} interpolates between two unequal expressions $\calK_\alpha^\mathrm{IR}(z) \ne \calK_\alpha^\mathrm{UV}(z)$ as the parameter $a$ varies. However, it actually does not depend on $a$ at all!

\section{Summary of analytical results} \label{SummaryAppendix}

In this Appendix, for the convenience of the reader, we have collected in a compact form all the analytical results we have obtained. We successively consider operator functions of four types: $\hat F^{-\mu}$, $\exp(-\tau\hat F)/\hat F^\mu$, $1/(\hat F^\mu + \lambda)$ and, finally, $\exp(-\tau\hat F)/(\hat F^\mu + \lambda)$. For each of them, we find a representation in the form of 1-, 2-, and 3-fold MB integrals for the basis $\bbB_\alpha\![f(F)| \sigma]$ \eqref{frakLtransformEq2}, complementary $\bbM_\alpha\![f(F)| m^2]$ \eqref{ComplInitEq}, and complete massive $\bbW_\alpha\![f(F)| \sigma, m^2]$ \eqref{BHtransform} kernels. In addition, we are considering objects of the fourth type $\bbL_\alpha\![f(F)]$, they can be obtained from $\bbB_\alpha$ by taking the coincidence limit $\sigma\to0$ or from $\bbM_\alpha$ by taking the massless limit $m^2\to0$, and therefore they depend only on the parameter $\alpha$ and the function $f$, but not on $\sigma$ and $m^2$. For each obtained kernel, we give its coincidence $\sigma\to0$ and massless $m^2\to0$ limits, as well as the limits $\tau\to0$ and $\lambda\to0$.

We use the following conventions: if $H_k(\bm{z})$ is an $N$-fold MB integral, then $h_k(\bm{s})$ is its Mellin image. If nothing is written above the limit arrow, this means that this limit holds for all values of the parameters $\alpha$ and $\mu$. If there is `rel' or `irrel' above the arrow, then the limit takes place only in the corresponding parameter region, clearly indicated immediately before the limit.

\paragraph{The operator complex power $\hat F^{-\mu}$}

\begin{align}
&\bbB_\alpha\!\big[F^{-\mu} \big| \sigma\big] = \frac{\Gamma\left(\alpha-\mu\right)}{(4\pi)^{d/2}\Gamma(\mu)} \left(\frac{\sigma}{2}\right)^{\mu-\alpha}, \label{CompPowFunctionsApp} \\
&\bbM_\alpha\!\big[F^{-\mu} \big| m^2\big] = \frac{\Gamma\left(\mu-\alpha\right)}{(4\pi)^{d/2}\Gamma(\mu)} m^{2(\alpha-\mu)}, \label{CompPowFunctions2App} \\
&\bbW_\alpha\!\big[F^{-\mu} \big| \sigma, m^2\big] = \frac{2m^{2(\alpha-\mu)}}{(4\pi)^{d/2} \Gamma(\mu)}\; \calK_{\alpha-\mu}\!\left(\sigma m^2/2 \right), \label{GreenIRregApp}
\end{align}
where $\calK_\alpha(z)$ is the Bessel--Clifford function of the second kind \eqref{BCMeelinBarnes2}-\eqref{BCMeelinBarnes}.

Limits: for relevant region $\Re(\alpha-\mu)>0$
\begin{equation} \label{LimitM2to0ConvApp}
\bbW_\alpha\!\big[ F^{-\mu} \big| \sigma, m^2\big] \xrightarrow[m^2\to0]{\mathrm{rel}} \bbB_\alpha\!\big[ F^{-\mu} \big| \sigma\big],
\end{equation}
for irrelevant region $\Re(\alpha-\mu)<0$
\begin{equation} \label{DivergenceRegionLimitApp}
\bbW_\alpha\!\big[ F^{-\mu} \big| \sigma, m^2\big] \xrightarrow[\sigma \to0]{\mathrm{irrel}} \bbM_\alpha\!\big[ F^{-\mu} \big| m^2\big].
\end{equation}

\paragraph{The operator function $\exp(-\tau\hat F^\nu)/\hat F^{\mu}$}

\begin{align}
&\bbB_\alpha\!\Big[ \frac{e^{-\tau F^\nu}}{F^\mu} \Big| \sigma\Big] = \frac{\tau^\frac{\mu-\alpha}{\nu}}{(4\pi)^{d/2}}\; \calE_{\nu,\alpha}^{(\mu)}\!\left(-\frac{\sigma}{2\tau^{1/\nu}}\right), \label{bbK_mn_from_bbK_0n} \\
&\bbM_\alpha\!\Big[ \frac{e^{-\tau F^\nu}}{F^\mu} \Big| m^2\Big] = \frac{\tau^\frac{\mu-\alpha}{\nu}}{(4\pi)^{d/2}}\; \tilde\calE_{\nu,\alpha}^{(\mu)}\!\left(m^2\tau^{1/\nu}\right), \label{bbMtildeE}
\end{align}
where
\begin{align}
&\calE_{\nu,\alpha}^{(\mu)}(-z) = \int\limits_C \frac{ds}{2\pi i}\, \frac{\Gamma(s)\Gamma\left(\frac{\alpha-s-\mu}{\nu}\right)}{\nu\Gamma(\alpha-s)}\, z^{-s}, \label{calKMellin} \\
&\tilde\calE_{\nu,\alpha}^{(\mu)}(z) = \int\limits_C \frac{ds}{2\pi i}\, \frac{\Gamma(s)\Gamma\left(\frac{s+\alpha-\mu}{\nu}\right)}{\nu\Gamma(s+\alpha)}\, z^{-s}. \label{calKMellin2}
\end{align}

\begin{align} 
&\bbL_\alpha\!\Big[ \frac{e^{-\tau F^\nu}}{F^\mu} \Big] = \frac{\Gamma\left(\frac{\alpha-\mu}{\nu}\right)}{(4\pi)^{d/2}\nu\Gamma(\alpha)} \tau^\frac{\mu-\alpha}{\nu}, \label{NulLKernellAppEq1} \\
&\bbW_\alpha\!\Big[ \frac{e^{-\tau F^\nu}}{F^\mu} \Big| \sigma, m^2\Big] = \frac{m^{2\alpha} \tau^{\mu/\nu}}{(4\pi)^{d/2}}\, \tilde H_2\!\Big(\frac{\sigma}{2\tau^{1/\nu}}, m^2\tau^{1/\nu}\Big), \label{E1App}
\end{align}
where
\begin{equation}
\tilde h_2(s_1, s_2) = \Gamma(s_1) \Gamma(s_2-\alpha) \frac{\Gamma\left(\frac{s_2-s_1-\mu}{\nu}\right)}{\nu\Gamma(s_2-s_1)}. \label{E3App}
\end{equation}

Limits:
\begin{align}
&\bbB_\alpha\!\Big[ \frac{e^{-\tau F^\nu}}{F^\mu} \Big| \sigma\Big]\xrightarrow[\tau\to0]{} \bbB_\alpha\!\big[ F^{-\mu} \big| \sigma\big], \label{Lim21Eq} \\
&\bbB_\alpha\!\Big[ \frac{e^{-\tau F^\nu}}{F^\mu} \Big| \sigma \Big]\xrightarrow[\sigma\to0]{} \bbL_\alpha\!\Big[ \frac{e^{-\tau F^\nu}}{F^\mu} \Big], \label{Lim22Eq} \\
&\bbW_\alpha\!\Big[\frac{e^{-\tau F^\nu}}{F^\mu} \Big| \sigma, m^2 \Big]\xrightarrow[\tau\to0]{} \bbW_\alpha\!\big[ F^{-\mu} \big| \sigma, m^2 \big], \label{LimTauTo0} \\
&\bbW_\alpha\!\Big[ \frac{e^{-\tau F^\nu}}{F^\mu} \Big| \sigma, m^2\Big]\xrightarrow[\sigma\to0]{} \bbM_\alpha\!\Big[ \frac{e^{-\tau F^\nu}}{F^\mu} \Big| m^2\Big]. \label{LimSigmaTo0}
\end{align}
For relevant region $\Re(\alpha-\mu)>0$
\begin{align}
&\bbM_\alpha\!\Big[ \tfrac{e^{-\tau F^\nu}}{F^\mu} \Big| m^2 \Big] \xrightarrow[m^2\to0]{\mathrm{rel}} \bbL_\alpha\!\Big[ \tfrac{e^{-\tau F^\nu}}{F^\mu} \Big], \\
&\bbW_\alpha\!\Big[ \frac{e^{-\tau F^\nu}}{F^\mu} \Big| \sigma, m^2\Big] \xrightarrow[m^2\to0]{\mathrm{rel}} \bbB_\alpha\!\Big[ \frac{e^{-\tau F^\nu}}{F^\mu} \Big| \sigma\Big]. \label{HuvLim}
\end{align}
For irrelevant region $\Re(\alpha-\mu)<0$
\begin{align}
&\bbM_\alpha\!\Big[ \tfrac{e^{-\tau F^\nu}}{F^\mu} \Big| m^2 \Big] \xrightarrow[\tau\to0]{\mathrm{irrel}} \bbM_\alpha\!\big[ F^{-\mu} \big| m^2 \big], \\
&\bbW_\alpha\!\Big[ \frac{e^{-\tau F^\nu}}{F^\mu} \Big| \sigma, m^2\Big] \xrightarrow[m^2\to0]{\mathrm{irrel}} \bbM_\alpha[ F^{-\mu}| m^2 ]. \label{HirLim}
\end{align}

\paragraph{The operator function $1/(\hat F^{\mu}+\lambda)$}

\begin{align}
&\bbB_\alpha\!\big[ (F^\mu + \lambda)^{-1} \big| \sigma\big] = \frac{\lambda^{\frac{\alpha}{\mu}-1}}{(4\pi)^{d/2}}\; \calG_{\mu,\alpha}\!\big(\sigma \lambda^{1/\mu}/2\big), \label{calGEq1} \\
&\bbM_\alpha\!\big[ (F^\mu + \lambda)^{-1} \big| m^2\big] = \frac{\lambda^{\frac{\alpha}{\mu}-1}}{(4\pi)^{d/2}}\; \tilde\calG_{\mu,\alpha}\!\big(m^2\lambda^{-1/\mu}\big), \label{calGEq2}
\end{align}
where
\begin{align}
&\calG_{\mu,\alpha}(z) = \int\limits_C \frac{ds}{2\pi i}\, \tfrac{\Gamma(s)}{\mu\Gamma(\alpha-s)}\,\Gamma\!\left(\tfrac{\alpha-s}{\mu}\right)\,\Gamma\!\left(1-\tfrac{\alpha-s}{\mu}\right)\, z^{-s}, \label{calGEq3} \\
&\tilde\calG_{\mu,\alpha}(z) = \int\limits_C \frac{ds}{2\pi i}\, \tfrac{\Gamma(s)}{\mu\Gamma(\alpha+s)}\,\Gamma\!\left(\tfrac{\alpha+s}{\mu}\right)\,\Gamma\!\left(1-\tfrac{\alpha+s}{\mu}\right)\, z^{-s}. \label{calGEq4}
\end{align}

\begin{align}
&\bbL_\alpha\!\big[ (F^\mu + \lambda)^{-1} \big] = \frac{\pi \lambda^{\frac{\alpha}{\mu}-1}}{(4\pi)^{d/2}\mu\Gamma(\alpha)\sin(\pi\alpha/\mu)}, \\
&\bbW_\alpha\!\big[ (F^\mu + \lambda)^{-1} \big| \sigma, m^2\big] = \frac{m^{2(\alpha-\mu)}}{(4\pi)^{d/2}}\, \tilde H_3\!\Big(\frac{\sigma m^2}{2}, \frac{\sigma\lambda^{1/\mu}}{2}\Big), \label{MlambdaEq3App}
\end{align}
where
\begin{equation}
\tilde h_3(s_1, s_2) = \Gamma(s_1+\mu-\alpha)\,\Gamma(s_1+s_2)\, \frac{\Gamma(\tfrac{s_2}{\mu})\,\Gamma(1-\tfrac{s_2}{\mu})}{\mu\Gamma(\mu-s_2)}. \label{MlambdaEq4App}
\end{equation}

Limits:
\begin{align}
&\bbM_\alpha\!\big[ (F^\mu + \lambda)^{-1} \big| m^2 \big] \xrightarrow[\lambda\to0]{} \bbM_\alpha\!\big[ F^{-\mu} \big| m^2 \big], \\
&\bbW_\alpha\!\big[ (F^\mu + \lambda)^{-1} \big| \sigma, m^2\big] \xrightarrow[\lambda\to0]{} \bbW_\alpha\!\big[ F^{-\mu} \big| \sigma, m^2\big]. \label{Hlambdato0Lim}
\end{align}
For relevant region $\Re(\alpha-\mu)>0$
\begin{align}
&\bbB_\alpha\!\big[ (F^\mu + \lambda)^{-1} \big| \sigma\big] \xrightarrow[\lambda\to0]{\mathrm{rel}} \bbB_\alpha\!\big[ F^{-\mu} \big| \sigma\big], \\
&\bbW_\alpha\!\big[ (F^\mu + \lambda)^{-1} \big| \sigma, m^2\big] \xrightarrow[\sigma\to0]{\mathrm{rel}} \bbB_\alpha\!\big[ F^{-\mu} \big| \sigma\big]. \label{UVsigmaHLim}
\end{align}
For irrelevant region $\Re(\alpha-\mu)<0$
\begin{align}
&\bbB_\alpha\!\big[ (F^\mu + \lambda)^{-1} \big| \sigma\big] \xrightarrow[\sigma\to0]{\mathrm{irrel}} \bbL_\alpha\!\big[ (F^\mu + \lambda)^{-1} \big], \label{LambdaCoLim} \\
&\bbW_\alpha\!\big[ (F^\mu + \lambda)^{-1} \big| \sigma, m^2\big] \xrightarrow[\sigma\to0]{\mathrm{irrel}} \bbM_\alpha\!\big[ (F^\mu + \lambda)^{-1} \big| m^2\big]. \label{IRsigmaHLim}
\end{align}
For $\Re(\alpha+\mu)>0$
\begin{align}
&\bbM_\alpha\!\big[ (F^\mu + \lambda)^{-1} \big| m^2 \big] \xrightarrow[m^2\to0]{\text{rel}} \bbL_\alpha\!\big[ (F^{\mu} + \lambda)^{-1} \big], \\
&\bbW_\alpha\!\big[ (F^\mu + \lambda)^{-1} \big| \sigma, m^2\big] \xrightarrow[m^2\to0]{\text{rel}} \bbB_\alpha\!\big[ (F^\mu+\lambda)^{-1} \big| \sigma\big]. \label{UVmHLim}
\end{align}

\paragraph{The operator function $\exp(-\tau\hat F^\nu)/(\hat F^{\mu}+\lambda)$}

\begin{align}
&\bbB_\alpha\!\Big[\frac{e^{-\tau F^\nu}}{F^\mu + \lambda} \Big| \sigma \Big] = \frac{\tau^\frac{\mu-\alpha}{\nu}}{(4\pi)^{d/2}}\; H_4\Big(\frac{\sigma}{2\tau^{1/\nu}}, \lambda^{1/\mu}\tau^{1/\nu}\Big), \label{H4expressionApp} \\
&\bbM_\alpha\!\Big[\frac{e^{-\tau F^\nu}}{F^\mu + \lambda} \Big| m^2 \Big] = \frac{\tau^\frac{\mu-\alpha}{\nu}}{(4\pi)^{d/2}}\; \tilde H_4\Big(m^2\tau^{1/\nu}, \lambda^{1/\mu}\tau^{1/\nu}\Big),
\end{align}
where
\begin{align}
&h_4(s_1, s_2) = \frac{\Gamma(s_1)\, \Gamma\!\big(\tfrac{s_2}{\mu}\big)\, \Gamma\!\big(1-\tfrac{s_2}{\mu}\big)}{\mu\nu\Gamma(\alpha-s_1)}\, \Gamma\!\left(\tfrac{s_2-s_1+\alpha-\mu}{\nu}\right), \label{H4MBApp}\\
&\tilde h_4(s_1, s_2) = \frac{\Gamma(s_1)\, \Gamma\!\big(\tfrac{s_2}{\mu}\big)\, \Gamma\!\big(1-\tfrac{s_2}{\mu}\big)}{\mu\nu\Gamma(\alpha+s_1)}\, \Gamma\!\left(\tfrac{s_2+s_1+\alpha-\mu}{\nu}\right).
\end{align}

\begin{equation}
\bbL_\alpha\!\Big[\frac{e^{-\tau F^\nu}}{F^\mu + \lambda} \Big] = \frac{\tau^\frac{\mu-\alpha}{\nu}}{(4\pi)^{d/2}\mu\nu\Gamma(\alpha)}\; \calL_{\nu,\alpha}^{(\mu)}\!\big( \lambda^{1/\mu}\tau^{1/\nu} \big),
\end{equation}
where
\begin{equation}
\calL_{\nu,\alpha}^{(\mu)}(z) = \int\limits_C \frac{ds}{2\pi i}\; \Gamma\!\left(\tfrac{s+\alpha-\mu}{\nu}\right)\, \Gamma\!\big(\tfrac{s}{\mu}\big)\, \Gamma\!\big(1-\tfrac{s}{\mu}\big)\, z^{-s}.
\end{equation}

\begin{equation}
\bbW_\alpha\!\Big[ \frac{e^{-\tau F^\nu}}{F^\mu+\lambda} \Big| \sigma, m^2 \Big] = \frac{m^{2\alpha}\tau^\frac{\mu}{\nu}}{(4\pi)^\frac{d}{2}}\; H_5\!\Big(\frac{\sigma}{2\tau^\frac{1}{\nu}}, m^2\tau^\frac{1}{\nu}, \lambda^\frac{1}{\mu}\tau^\frac{1}{\nu}\Big),
\end{equation}
where
\begin{equation}
h_5(s_1, s_2, s_3) = \tfrac{\Gamma(s_1) \Gamma(s_2-\alpha) \Gamma(\tfrac{s_3}{\mu}) \Gamma(1-\tfrac{s_3}{\mu}) \Gamma(\tfrac{s_2-s_1+s_3-\mu}{\nu})}{\mu\nu\Gamma(s_2-s_1)}.
\end{equation}

Limits:
\begin{align}
&\bbB_\alpha\!\Big[\frac{e^{-\tau F^\nu}}{F^\mu + \lambda} \Big| \sigma \Big] \xrightarrow[\sigma\to0]{} \bbL_\alpha\!\Big[\frac{e^{-\tau F^\nu}}{F^\mu + \lambda} \Big], \\
&\bbB_\alpha\!\Big[\frac{e^{-\tau F^\nu}}{F^\mu + \lambda} \Big| \sigma \Big] \xrightarrow[\tau\to0]{} \bbB_\alpha\!\big[(F^\mu + \lambda)^{-1} \big| \sigma \big], \\
&\bbM_\alpha\!\Big[\frac{e^{-\tau F^\nu}}{F^\mu + \lambda} \Big| m^2 \Big] \xrightarrow[\lambda\to0]{} \bbM_\alpha\!\Big[\frac{e^{-\tau F^\nu}}{F^\mu} \Big| m^2 \Big].
\end{align}
For relevant region $\Re(\alpha-\mu)>0$
\begin{align}
&\bbB_\alpha\!\Big[\frac{e^{-\tau F^\nu}}{F^\mu + \lambda} \Big| \sigma \Big] \xrightarrow[\lambda\to0]{\mathrm{rel}} \bbB_\alpha\!\Big[\frac{e^{-\tau F^\nu}}{F^\mu} \Big| \sigma \Big], \\
&\bbM_\alpha\!\Big[\frac{e^{-\tau F^\nu}}{F^\mu + \lambda} \Big| m^2 \Big] \xrightarrow[\tau\to0]{\mathrm{rel}} \bbL_\alpha\!\Big[\frac{e^{-\tau F^\nu}}{F^\mu} \Big], \\
&\bbL_\alpha\!\Big[\frac{e^{-\tau F^\nu}}{F^\mu + \lambda} \Big] \xrightarrow[\lambda\to0]{\mathrm{rel}} \bbL_\alpha\!\Big[\frac{e^{-\tau F^\nu}}{F^\mu} \Big].
\end{align}
For irrelevant region $\Re(\alpha-\mu)<0$
\begin{align}
&\bbB_\alpha\!\Big[\frac{e^{-\tau F^\nu}}{F^\mu + \lambda} \Big| \sigma \Big] \xrightarrow[\lambda\to0]{\mathrm{irrel}} \bbL_\alpha\!\big[ (F^\mu + \lambda)^{-1} \big], \\
&\bbM_\alpha\!\Big[\frac{e^{-\tau F^\nu}}{F^\mu + \lambda} \Big| m^2 \Big] \xrightarrow[\tau\to0]{\mathrm{irrel}} \bbM_\alpha\!\big[ (F^\mu + \lambda)^{-1} \big| m^2 \big], \\
&\bbL_\alpha\!\Big[\frac{e^{-\tau F^\nu}}{F^\mu + \lambda} \Big] \xrightarrow[\tau\to0]{\mathrm{irrel}} \bbL_\alpha\!\big[ (F^\mu + \lambda)^{-1} \big].
\end{align}

\bibliography{Wachowski2512}
\end{document}